\documentclass[%
 reprint,
 amsmath,amssymb,
 aps,
]{revtex4-2}

\usepackage{graphicx}% Include figure files
\usepackage{dcolumn}% Align table columns on decimal point
\usepackage{bm}% bold math
\usepackage{subcaption}
\usepackage{caption}

\usepackage{siunitx}
\usepackage{comment}
\usepackage{color}
\usepackage[dvipsnames]{xcolor}
\usepackage{tikz}
\usepackage[margin=1in]{geometry}

\begin{document}

\preprint{APS/123-QED}

\title{Rheological Insights from the Oscillation Dynamics of Viscoelastic Sessile Drops}% Force line breaks with \\

\author{Peyman Rostami}
 \email{rostami@ipfdd.de}
\affiliation{ 
Leibniz-Institut f\"ur Polymerforschung  Dresden e.V., Hohe Str. 6, 01069 Dresden, Germany  %\\This line brforced with \textbackslash\textbackslash
}%Lines break automatically or can be forced with \\

\author{Alfonso A. Castrej\'on-Pita}%
 \email{alfonso.castrejon-pita@eng.ox.ac.uk}
\affiliation{%
 Department of Engineering Science, University of Oxford, Oxford OX1 3PJ, United Kingdom
}%

\author{Günter K. Auernhammer }
\email{auernhammer@ipfdd.de}
\affiliation{ Leibniz-Institut  f\"ur Polymerforschung Dresden e.V.,  Hohe Str. 6 ,  01069 Dresden,  Germany %\\This line break forced with \textbackslash\textbackslash
}


\date{\today}% It is always \today, today,

\begin{abstract}
This study investigates the oscillation behavior of a sessile drop placed on a hydrophobic substrate subjected to vertical vibrations with varying frequencies and amplitudes.
We examined the responses of both Newtonian and viscoelastic drops. For viscoelastic samples, image analysis techniques were employed to correlate the drop dynamics with the rheological properties of the material.
Overall, we demonstrate that this drop-based method allows for oscillatory shear experiments at frequencies that are difficult to access using conventional rheometers.
The results reveal that the essential features of the drop response can be explained by the ratio of two characteristic time scales: the internal polymer relaxation time ($t_{p}$) and the external forcing time scale ($1/f$).
This ratio defines the Deborah number ($De$).
When the two time scales are comparable ($De \approx 1$), viscous dissipation dominates, which is observed in Lissajous curves and the drop's profile.
At very low Deborah numbers ($De \ll 1$), the drop behaves like a Newtonian fluid (having a peak around natural frequency of the drop), while at high Deborah numbers ($De \gg 1$), it exhibits an elastic response.
Furthermore, we show that increasing the applied deformation drives the system into the nonlinear viscoelastic regime.
In this regime, unlike traditional rheology measurements, we observe the presence of $even$ and $odd$ harmonics in the drop response.
This is attributed to the inherent geometric asymmetry of the drop setup, which breaks the symmetric assumptions typically present in standard rheological techniques.

\end{abstract}

\maketitle

%\tableofcontents

\section{\label{sec:level1} Introduction}

The oscillation of sessile drops on flat substrates plays a crucial role in various applications, including energy harvesting \cite{lee2015energy}, inkjet printing \cite{lohse2022fundamental}, and other processes involving substrate vibration.
This vibration can enhance the mobility of the drop on the surface and, in some cases, lead to detachment from the substrate or breakage into smaller drops \cite{yarin2002acoustically, Mettu_2010aa, Manor_2011aa,Montes-Ruiz-Cabello_2011aa}.
The oscillation of a freely levitating drop is also used to measure the surface tension \cite{arcenegui2019simple} and viscosity of the liquid \cite{myrvold1998surface}.
Pioneering work by Lord Rayleigh \cite{rayleigh1879capillary} assumed an inviscid condition (e.g., a water drop) and derived the natural frequency: ($ \omega _{n}^{2}=n(n-1)(n+2)\frac{4\pi }{3}\frac{\sigma }{m}$).
Here, $\sigma$ is the surface tension and $m$ is the drop mass.
Later, the effects of the drop's shape and the surrounding fluid were added to this theory by Lamb \cite{lamb1924hydrodynamics}.
If the drop oscillates at or near its natural frequency ($\omega_n$), the oscillation amplitude can become significantly amplified, potentially leading to secondary drop breakup.
It was shown that increasing drop viscosity leads to a decrease in the drop deformation amplitude as well as the natural frequency \cite{lamb1924hydrodynamics, bostwick2014dynamics}.
This simple prediction remains valid in various scenarios, such as a drop immersed in another liquid \cite{trinh1982experimental}, a drop on a non-wetting substrate \cite{brunet2011star}, and a drop on a wetting substrate \cite{noblin2009ratchetlike}.

For sessile drops, the situation becomes more complex due to the presence of the substrate~\cite{chang2013substrate}.  
One of the critical parameters is the drop's contact angle ($\theta$), which indirectly controls the drop free surface area for a fixed volume.  
It has been shown that increasing the contact angle, increases the surface energy restoring pressure and decreases the natural frequency, which approaches the asymptotic solution for a free drop (i.e., when $\theta = \SI{180}{\degree}$)~\cite{bostwick2014dynamics}.  

Another relevant parameter is gravity, which can be characterized by the Bond number (the ratio between gravitational to surface tension forces): 

\begin{equation}
Bo = \frac{\rho g R^{2}}{\sigma}
\label{Eq-Bond}
\end{equation}

As the Bond number increases (i.e., for larger drops of the same liquid), the natural frequency decreases~\cite{basaran1994nonlinear,sakakeeny2021model}. In the present study, we have kept the Bond number below 1 to remain in the surface-tension-dominated regime.

Another point we discuss here is contact line mobility.  
There are two extreme cases:  
in the first, the contact line can freely move over the substrate (e.g., drops on liquid-infused surfaces).  
In this case, the contact angle hysteresis is assumed to be zero, and the contact angle remains at its equilibrium value.  
This behavior is referred to as the "free contact line" (FCL) mode.  
At the other extreme, the contact line is pinned, and the contact angle varies during oscillation due to nonzero contact angle hysteresis.  
This is known as the "pinned contact line" (PCL) mode.  
It has been shown that, in general, the PCL mode exhibits a higher natural frequency~\cite{bostwick2014dynamics} and dissipates more energy.
In the present study, we have kept the contact line pinned to the surface.

There have been several reports on how to measure the physical properties of liquids based on their response to oscillations~\cite{egry2005oscillating,arcenegui2019simple,singha2021oscillating}.  
In most cases, surface tension is measured using the vibration of a levitated drop~\cite{trinh1988acoustic,argyri2023contact}.  
By applying a short mechanical impulse to a sessile drop on a superamphiphobic substrate, V.~C.~Harrold and J.~S.~Sharp~\cite{harrold2016optovibrometry} were able to measure the surface tension and viscosity of aqueous glycerin solutions.  
Another attempt to measure viscosity from the oscillation of a sessile drop was made by P.~H.~Steen and collaborators~\cite{chang2015dynamics,bostwick2016dynamics}.  
They solved the damped-driven oscillator equation (Eq.~\ref{Eq-damped-drivenoscillator}):

\begin{equation}
 m\ddot{h}+c\dot{h}+kh=F_{0}\sin(\omega t)
\label{Eq-damped-drivenoscillator}
\end{equation}

where $h$ is the drop height, $m$ is the drop inertia ($\sim$ mass $ \sim \rho R^{3}$), $c$ is the bulk dissipation term related to the viscosity ($\sim \eta R$), and $k$ represents the capillary force (spring constant, $\sim \sigma$).  
The source term $F_{0}\sin(\omega t)$ accounts for the external force acting on the substrate and the drop.  
By solving this equation and measuring the phase shift, it is possible to determine the relationship between the viscosity and the drop's contact angle at different frequencies~\cite{chang2015dynamics,bostwick2016dynamics}.

Most studies on drop oscillations have focused on Newtonian liquids, even though many applications involve more complex fluids.
There have been a few attempts to study more complex drops. 
Sharp et al., investigated the rheological properties of PAA polymer solutions by analyzing the surface waves of drops subjected to short mechanical impulses~\cite{harrold2016rheological}. 
A few theoretical and numerical studies have been conducted on viscoelastic levitating drops \cite{tamim2021oscillations, zrnic2024weakly}.
These results are limited to small amplitude and impulsive deformations.
In contrast if we consider continuous periodic deformation, the drop undergoes oscillatory strain and stress.
With these experiments, we can measure rheological properties of the drop that cannot be obtained from simple flow curves. 
In polymer physics, when the material response is proportional with the applied force, we are in linear viscoelastic regime (LVR).
If we perform oscillatory shear measurements (applying a small periodic deformation), the response remains in the LVR, and the method is known as "Small Amplitude Oscillatory Shear” (SAOS).
By increasing the applied force (or deformation) amplitude, the material response becomes non linear which is the hint of reaching the non-linear viscoelastic regime\cite{ferry1980viscoelastic,pipkin2012lectures}.  
For larger amplitudes, the liquid exhibits a nonlinear response, and this mode is called "Large Amplitude Oscillatory Shear" (LAOS)\cite{ewoldt2008new,hyun2011review}.  

For small amplitude oscillations, the measured stress only contains the first harmonic of the driving frequency, i.e., the driving frequency itself.
By increasing the oscillation amplitude, the response (i.e., stress) becomes nonlinear and higher harmonics appear, as described in Eq.~\ref{Eq-Higher-Order}:

\begin{equation}
  \tau(t) = \sum_{\substack{n=1\\n\ \text{odd}}}^{\infty} \tau_{n} \sin(n\omega t + \delta_{n}),
  \label{Eq-Higher-Order}
\end{equation}
 
The appearance of odd harmonics arises directly from the symmetry of oscillatory shear, meaning the system is symmetric under strain reversal.
To fulfill this condition, only odd harmonics are present in the response.
The ratio between higher harmonics amplitudes to the first amplitude ($\tau_{3}/\tau_{1}$, $\tau_{5}/\tau_{1}$), indicates the level of non-linearity. 
A more sensitive implementation for detecting and analyzing these nonlinearities is known as "Fourier Transform Rheology" (FT-Rheology) \cite{wilhelm2002fourier,hyun2009establishing}.
Furthermore, by plotting the strain ($\epsilon = \frac{h - h_{0}}{h_{0}}$), where $h$ and ${h_{0}}$ are drop height and average drop height or the strain rate ($ \dot{\epsilon}$) as a function of stress ($\tau$), we get the curves known as Lissajous curves.
These curves are closed loops formed by plotting two periodic signals against each other.
These Lissajous curves are commonly used to measure the rheological parameters such as storage modulus ($G'$) and loss modulus ($G''$) \cite{philippoff1964dynamic}.

In conventional rheometers, the motor applies deformation (either strain or strain rate), and a torque transducer measures the resulting torque and, subsequently, the stress.
In the case of "drop measurements", the substrate oscillation and the drop response can be directly measured using high-speed imaging techniques.
For these experiments, we have kept the drop size in the range of low Bond numbers ($Bo < 1$) and the contact line pinned. 
From the recorded data, the strain can be readily calculated, while the stress must be estimated.
The relevant stress components include inertial, viscous, elastic and capillary stresses.
The main objective of the present study is to demonstrate the potential of this new technique for measuring the rheological properties of complex liquids and to lay the groundwork for further advancements in the field.
This method may be particularly useful for measuring the rheological properties of expensive materials, as it requires only a small drop of liquid.
Additionally, it can access a broader range of frequencies compared to commercial rheometers.

\section{\label{sec:Material} Materials and Methods:}

We have used water and glycerin (Sigma-Aldrich Co., purity 99\%) as the inviscid Newtonian and viscous Newtonian liquids.
For the viscoelastic drops, we prepared water–PEO (Sigma-Aldrich Co.) solutions with various molar masses and weight concentrations.  
Here we just reported the relevant parameters for the current work.  
To measure the polymer relaxation time ($t_p$), we performed flow curve measurements using a commercial rheometer (MCR 502, Anton Paar GmbH) and measured the viscosity ($\eta$) as a function of shear rate ($\dot{\gamma}$).  
The viscosity data were fitted to the Cross fluid model (Eq.~\ref{Cross-model})~\cite{cross1965rheology}.  
The detailed physical properties and preparation method of these liquids are reported in our previous publications~\cite{rostami2024spreading,rostami2025coalescence}.

\begin{equation}
\eta=\frac{\eta_{0}-\eta_{\infty}}{1+(t_{p}\dot{\gamma})^{m}}+\eta_{0}
  \label{Cross-model}
\end{equation}

By considering the viscosity at infinite shear rate ($\eta_{\infty}$) as the viscosity of the base solvent (i.e., water), and the zero-shear-rate viscosity ($\eta_0$) as the viscosity at low shear rates $\dot{\gamma} =0.1$, we fitted the polymer relaxation time.  
All relevant parameters are listed in Table ~\ref{Table-Properties}.  
For all samples, concentration is in term of weight percentage, and $k$ denotes units of (\si{\kilo\gram\per\mol}).

\begin{table}
\caption{\label{Table-Properties}Composition of operating fluids, the molar masses (as given by the supplier, $k$ = \si{\kilo \gram \per \mol}), the zero-shear viscosity $\eta_{0}$  (at \SI{23}{\celsius} and $\dot{\gamma }= 0.1 (\frac{1}{s})$), the polymer relaxation time $t_{p}$ (s). }
\begin{ruledtabular}
\begin{tabular}{lll}
Sample & $\eta_{0}$ (\SI{}{\milli\pascal\cdot\second}) & $t _{p}$ (s)\\
\hline

Glycerin &  $900\pm 0.5$  &  0 \\
 PEO ($3\%, 300k$)   &  $101\pm 0.5$  &  0.0002186   \\
 PEO ($3\%, 1000k$)  &  $3340\pm 0.5$  &  0.0943   \\
 PEO ($1\%, 4000k$) &  $952\pm 0.5$  &  1.16  \\
 PEO ($0.5\%, 8000k$) &  $2450\pm 0.5$  &  2.31  \\
 
\end{tabular}
\end{ruledtabular}
\end{table}

\begin{figure}
\includegraphics[width=0.85\linewidth]{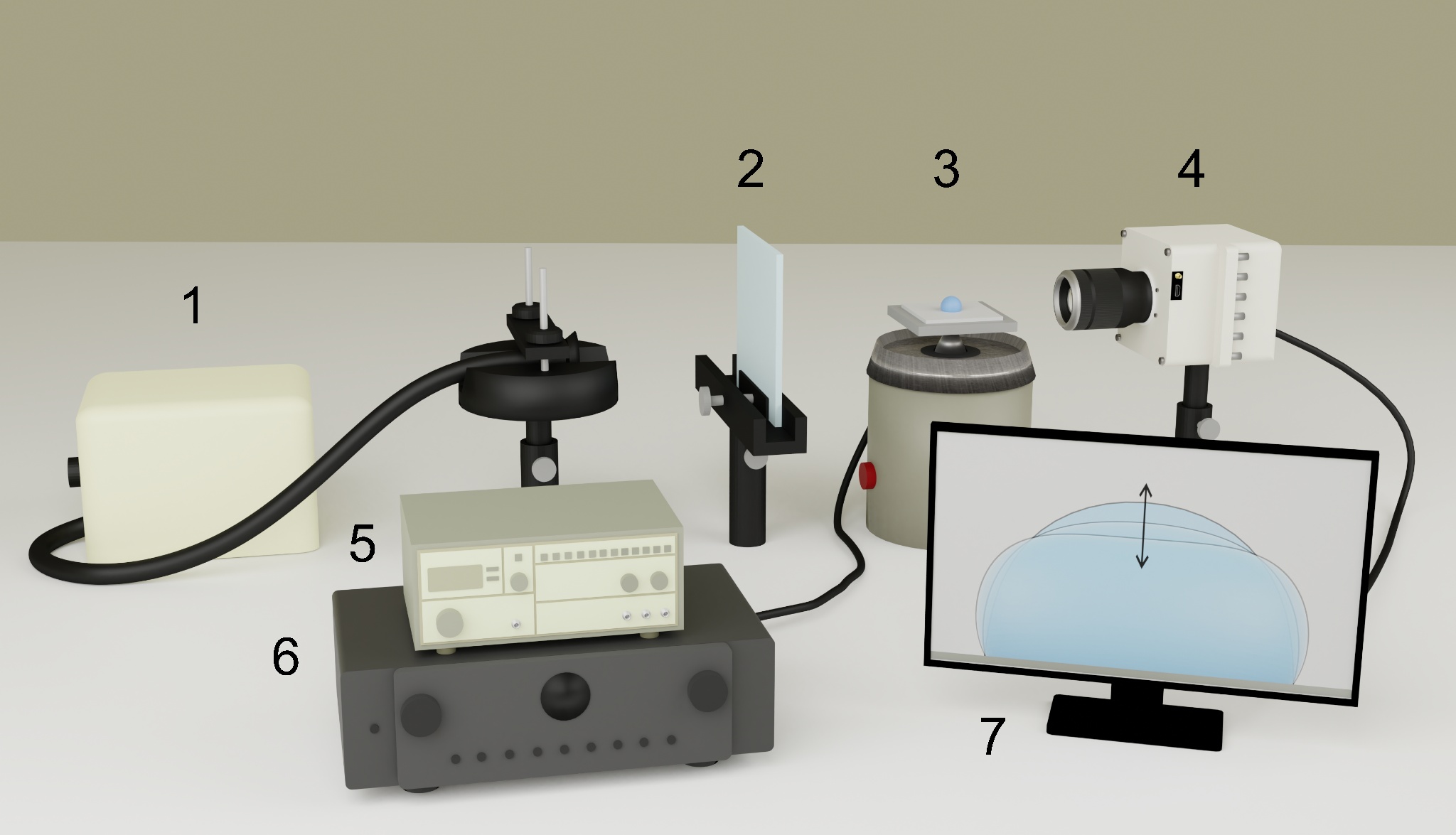}% Here is how to import EPS art
\caption{\label{Experimental-Set-Up} A sketch of drop oscillating experimental set-up, which has 1) LED 2) diffuser sheet 3) magnetic shaker 4) high speed camera 5) the signal generator 6) amplifier and 7) computer to record the images. }
\end{figure}

The experimental setup consists of an electromagnetic shaker (LDS V201, Brüel and Kjær).  
The shaker is driven by a Marantz P60 audio amplifier, which amplifies the input signal generated by a function generator (TTi TG5011, Thurlby Thandar Instruments), see Fig.~\ref{Experimental-Set-Up}.  
The substrates used were silicon wafers (Siegert Wafer, Germany), cut into \SI{1}{\square\centi\meter} pieces and cleaned using standard solvent cleaning procedures.
The substrates are submerged in acetone, ethanol, and water, and placed in an ultrasonic bath for 15 minutes in each solvent.
The substrates were then silanized with 1H,1H,2H,2H-perfluorooctyltriethoxysilane (Sigma-Aldrich, 97\%).
The advancing contact angle on these surfaces is approximately $\theta_{\text{adv}} \approx$ \SI{115}{\degree}.
A detailed description of the substrate preparation can be found in our previous publication \cite{rostami2023dynamic}.

To capture the drop and its oscillation, we used a high-speed camera (Phantom VEO 1310) in combination with an LED light source and a diffuser sheet (Fig.~\ref{Experimental-Set-Up}).  
We analyzed the recorded images using an in-house MATLAB script.  
These image processing steps allow us to measure the drop height, shape and the motion of the substrate as functions of time.  
A time sequence of the drop and the corresponding analyzed signals for the substrate motion and drop height are presented in the next sections.

\section{\label{sec:Results} Results:}
\subsection{Newtonian case}

\begin{figure}
\includegraphics[width=0.85\linewidth]{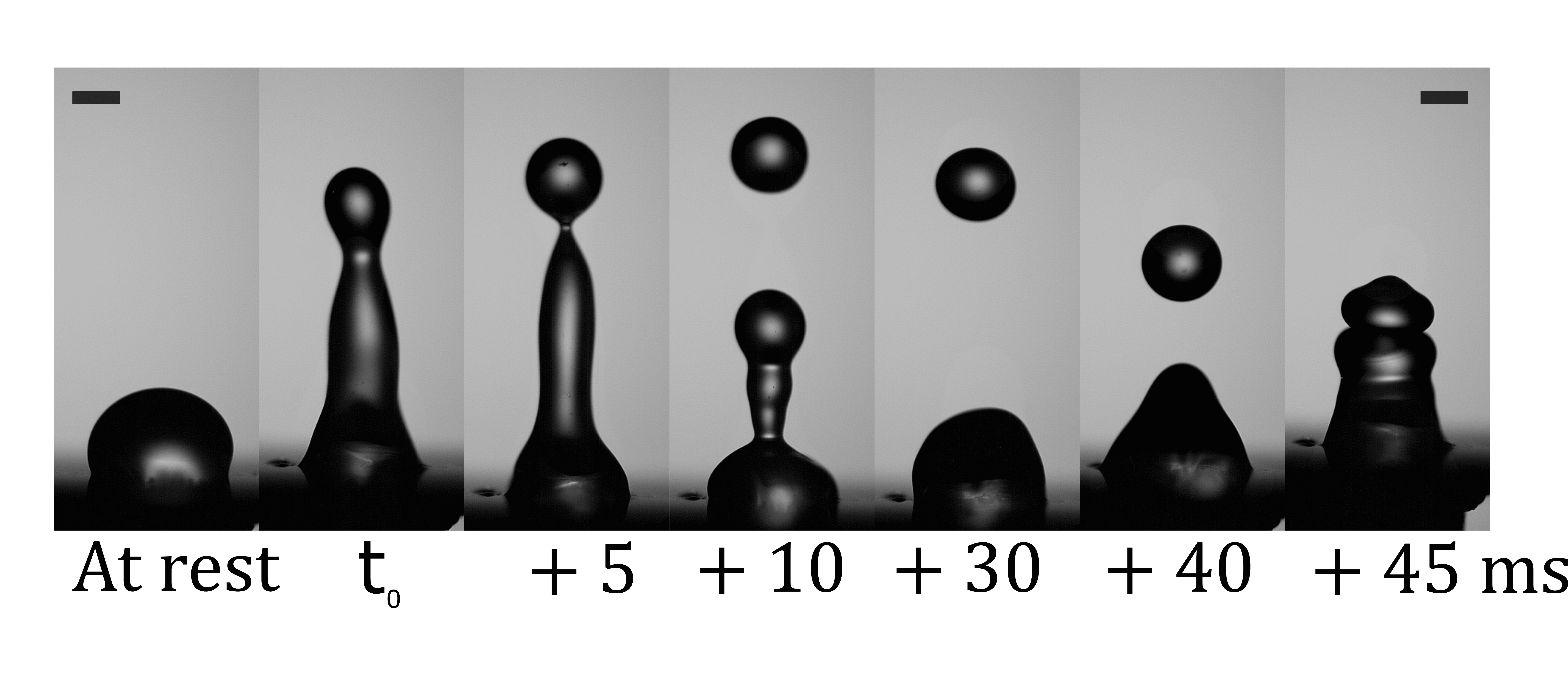}% Here is how to import EPS art
\caption{\label{Water-Frequency-Sweep} Snapshots of the water drop oscillation at \SI{45}{\hertz} with an amplitude of \SI{0.5}{\milli\meter} are shown. 
At this critical frequency, the onset of instability occurs, and both drop breakup and coalescence are visible. These breakups and coalescences occur periodically.
The first snapshot shows the drop at rest. The scale bar corresponds to \SI{1}{\milli\meter}.
}
\end{figure}

As mentioned in the introduction, most of published experiments have been performed on water drop oscillation.
In most cases, the focus has been on identifying the critical frequency (natural frequency) and observing the ejection of secondary drops \cite{james2003vibration,noblin2004vibrated}.
The relationship between oscillation modes and the driving frequency has also received significant attention \cite{bostwick2014dynamics,ding2024parametric}.
We performed experiments on a Newtonian drop to compare the results with previously published data and to use this as groundwork for the viscoelastic cases. 
To illustrate this behavior, a water drop was placed on a hydrophobic substrate (contact angle $\theta = $\SI{105}{\degree}) and subjected to a constant oscillation amplitude ($A = \SI{0.5}{\milli\meter}$).
A frequency sweep was performed from \SI{10}{\hertz} to \SI{100}{\hertz} (continuous frequency sweep). 
Around the critical frequency ($f_{\text{crit}} \approx \SI{45}{\hertz}$), instability occurred, leading to drop breakup.
Fig. ~\ref{Water-Frequency-Sweep} shows the drop at rest in contrast with the unstable drop at \SI{45}{\hertz}.
In this case, the applied acceleration due to substrate oscillation is well above gravitational acceleration ($\Gamma = \frac{A\omega^2}{g} \approx 4$).

Another way to determine the critical frequency is by plotting the deformation as a function of the driving frequency at a fixed amplitude (stepped frequency sweep). 
To quantify the deformation, we measured the strain ($\epsilon$) and plotted it against the driving frequency for a fixed amplitude ($A = \SI{0.1}{\milli\meter}$) and a fixed drop size ($h_{0} \approx \SI{2.5}{\milli\meter}$).
The main reason for choosing a small amplitude was to ensure that all assumptions of the linear regime remain valid:

\begin{itemize}
    \item  Small amplitude oscillations.
    \item Keeping the contact line pinned.
    \item No significant droplet detachment or breakup.
    
    \end{itemize}

\begin{figure}
\includegraphics[width=0.85\linewidth]{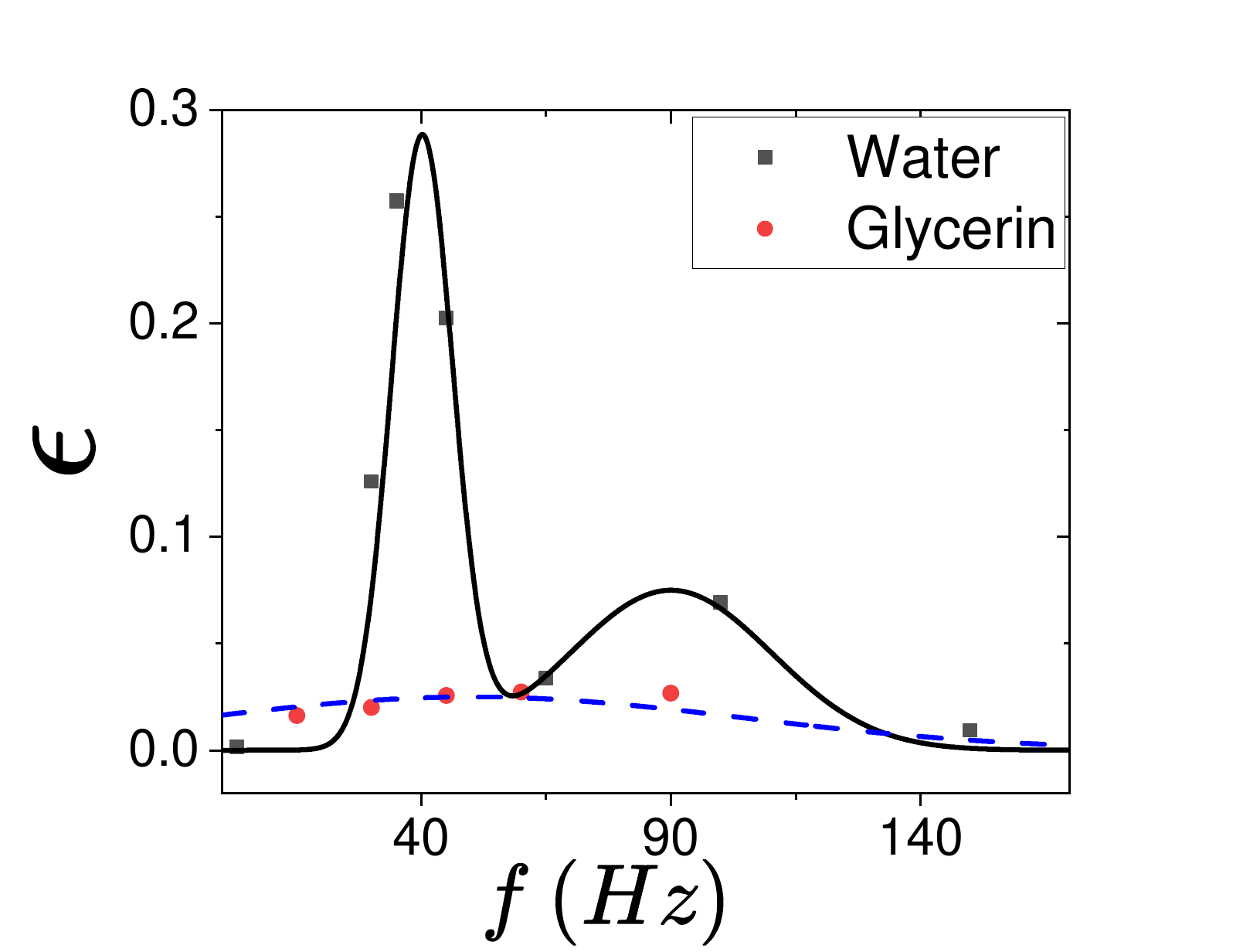}% Here is how to import EPS art
\caption{\label{Strain-Frequency} The applied strain as a function of frequency is shown for a water drop (black squares) and pure glycerin (red circles). Peaks corresponding to the water case are visible. The solid black line and the dashed blue line are shown as guides to the eye for water and glycerin.}
\end{figure}

When we measure the strain and plot it against the driving frequency, the results for water drops reveals two distinct peaks, Fig. \ref{Strain-Frequency} (black squares). 
The dominant peak appears around \SI{40}{\hertz}, which is close to the critical frequency identified in the continuous frequency sweep experiment (Fig. \ref{Water-Frequency-Sweep}).
The second peak which is at \SI{90}{\hertz} which is roughly 2 times of eigenfrequency.

Next we test a highly viscous drop.
We performed the same experiments using pure glycerin drops ($\eta = $ \SI{900}{\milli\pascal\second}).
The measured strains as a function of frequency are plotted in Fig. \ref{Strain-Frequency} (red circles).
In this case, no peak is observed in the strain, which can be attributed to the high viscosity of the drop.
To explain this behavior, we refer to the governing equation for a damped, driven oscillator (Eq\ref{Eq-damped-drivenoscillator}).
The steady-state solution of damped driven oscillator is $h(t)=A_{1}\sin(\omega t+\delta)$, where $A_{1}$ is a function of ratio between forcing frequency and natural frequency, as well as damping ratio ($\zeta$). 
The damping constant is given by $ \tfrac{c}{2m} $, and the eigenfrequency of the system is $ \sqrt{\tfrac{k}{m}} $.
The ratio of these two quantities defines the damping ratio ($ \zeta = \tfrac{c}{2\sqrt{km}} = \tfrac{\eta}{2\sqrt{\rho R \sigma}} $), which characterizes the system’s response under different conditions (i.e., different liquid properties).
Using this simplified approximation, the damping ratio is $\zeta_{\text{vis}} > 1$ for glycerin drops and $ \zeta_{\text{inv}} \ll  1$ for water drops.
For $\zeta > 1$ the system is overdamped, and the oscillatory behavior is suppressed.
In contrast, for $0 < \zeta < 1$, the system is under damped, and the drop exhibits oscillations with varying amplitudes.

We performed frequency sweep measurements on the glycerin drop. Within the accessible frequency and amplitude range of our shaker, no instability or drop breakup was observed.
To confirm that the oscillation amplitude does not play a significant role in viscous drop oscillations, we present an example of a glycerin drop placed on a vertically oscillating substrate at a driving frequency of \SI{45}{\hertz} and an amplitude of $A = \SI{0.9}{\milli\meter}$.

\begin{equation}
 y_{sub}=0.9\sin( 2\: \pi \:  45 \: t)
 \label{Eq-Substrate}
\end{equation}

\begin{figure}[htbp]
  \centering

  % Subfigure A: made of one image
  \begin{subfigure}[b]{\linewidth}
    \centering
    \includegraphics[width=0.85\textwidth]{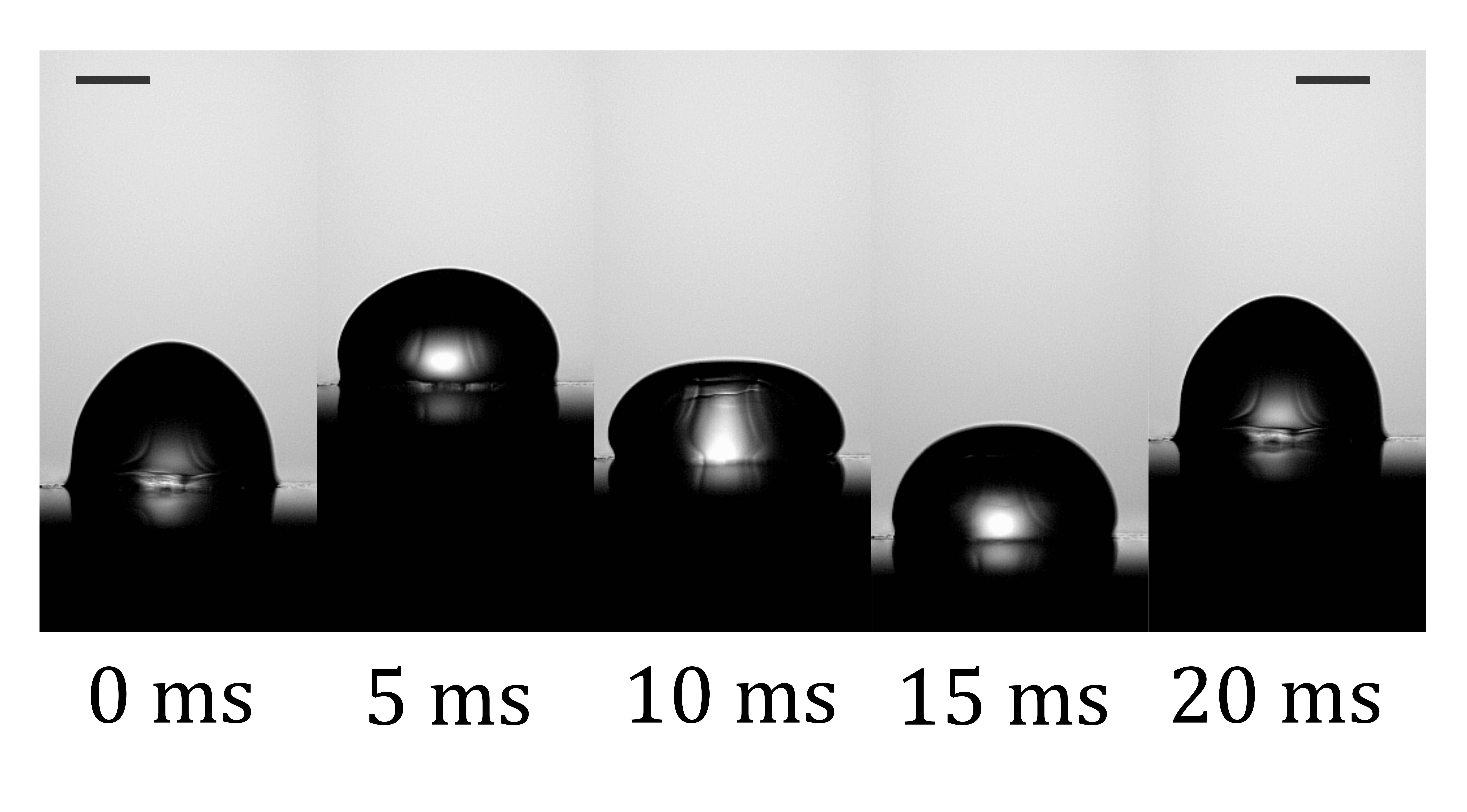}
    %\caption*{(A)}
  \end{subfigure}
  
  \vspace{1em}  % optional vertical space between subfigures

  % Subfigure B: made of two stacked images
  \begin{subfigure}[b]{\linewidth}
    \centering
    \includegraphics[width=0.85\linewidth]{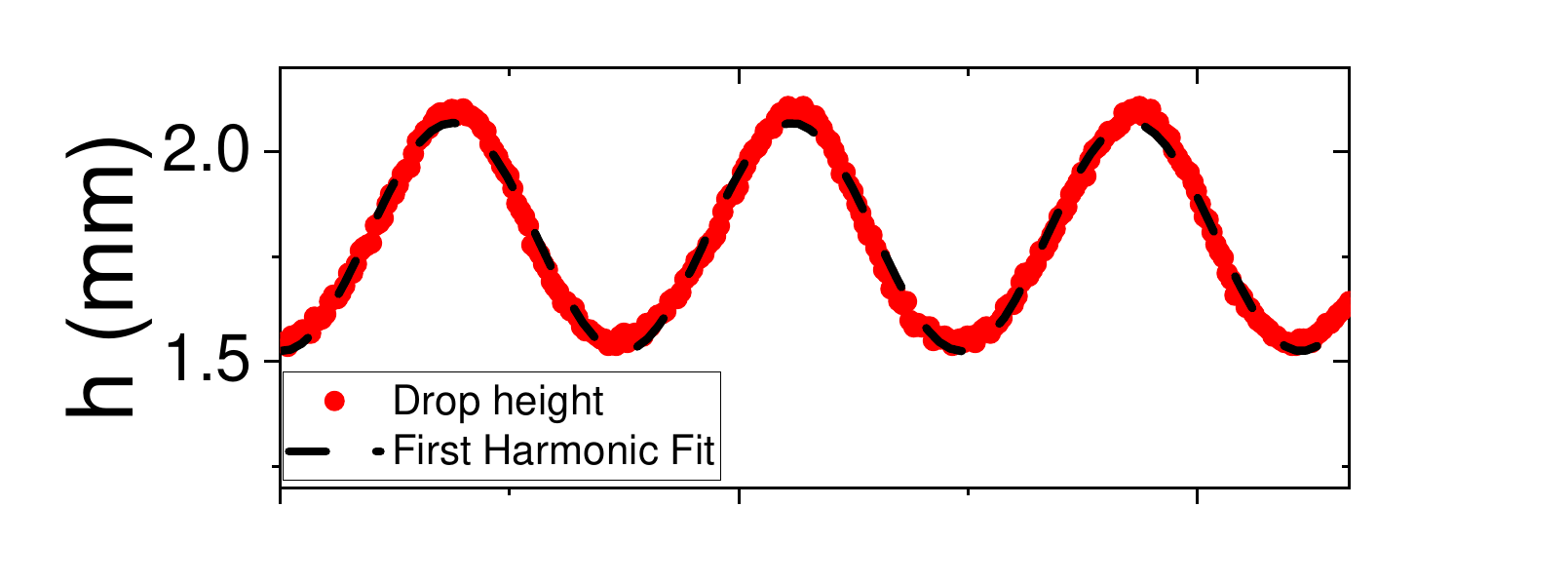}
    \includegraphics[width=0.85\linewidth]{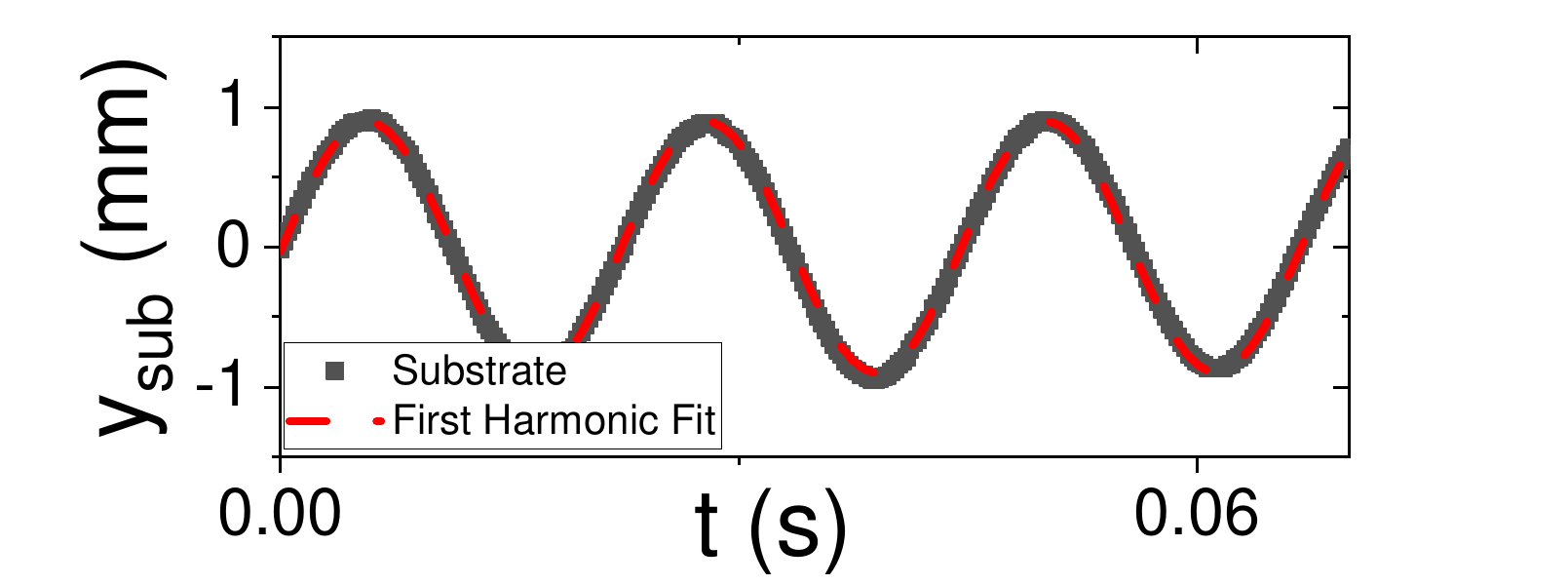}
    %\caption*{(B)}
  \end{subfigure}

  \caption{Top: The snapshot of an oscillating glycerin drop on a vertically oscillating substrate at \SI{45}{\hertz}.
Bottom: The measured values for the position of the substrate (black square) and the height of the drop (red circle) as function of time. The fitted sinusoidal curves are Eq.~\ref{Eq-Substrate} for the substrate and Eq.~\ref{Eq-Drop} for the drop height.}
  \label{Glycerin-Oscillation}
\end{figure}

In response, the drop height oscillates at the same frequency but with a phase shift ($\delta \approx \tfrac{\pi}{2}$; see Eq.\ref{Eq-Drop}).
A phase shift of $\delta \approx \tfrac{\pi}{2}$ indicates that the system is very close to the eigenfrequency of the drop, where deformation should be maximized.
In practice, however, the ratio of the drop’s oscillation amplitude to the forcing amplitude ($\tfrac{A_1}{A} = 0.3$) shows that viscous dissipation is dominant.
The time-dependent positions of the substrate and the drop height are shown in the lower part of Fig.\ref{Glycerin-Oscillation}.
The fitted sine functions for the substrate ($y_{\text{sub}}$) and the drop height ($h$) are given by Eq.\ref{Eq-Substrate} and .\ref{Eq-Drop}, respectively.”

\begin{equation}
h=1.8+0.27\sin( 2\: \pi \:  45 \: t- 0.51\pi)
 \label{Eq-Drop}
\end{equation}

In conclusion of this part, the oscillation of a Newtonian sessile drop on a vertically vibrating substrate follows the steady-state solution of a damped driven oscillator.
In this solution ($h(t) = A_{1}\sin(\omega t + \delta)$), the phase lag depends on the driving frequency and viscosity.
The phase lag can be calculated by: $\delta=\arctan(\frac{2\zeta \omega \omega_{n}}{\omega_{n}^{2}-\omega^{2}})$.
Near the eigenfrequency, the phase lag approaches \SI{90}{\degree}, and the deformation is maximized for under damped cases ($\zeta < 1$).
At lower and higher frequencies, the phase lag varies between \SI{0}{\degree} and \SI{180}{\degree}.

\subsection{Linear viscoelastic regime}

In this section, we present experiments performed on polymer solutions, as listed in Table~\ref{Table-Properties}.
When the applied force is small, the drop remains in the linear viscoelastic regime, and the deformation is also small ($\epsilon < $ 10\%).
In this regime, the drop responds at the same frequency as the vertical driving frequency of the substrate, analogous to the behavior observed in viscous drops.

\begin{figure}[htbp]
  \centering

  % Subfigure A: made of two vertically stacked images
  \begin{subfigure}[b]{\linewidth}
    \centering
    \includegraphics[width=0.85\linewidth]{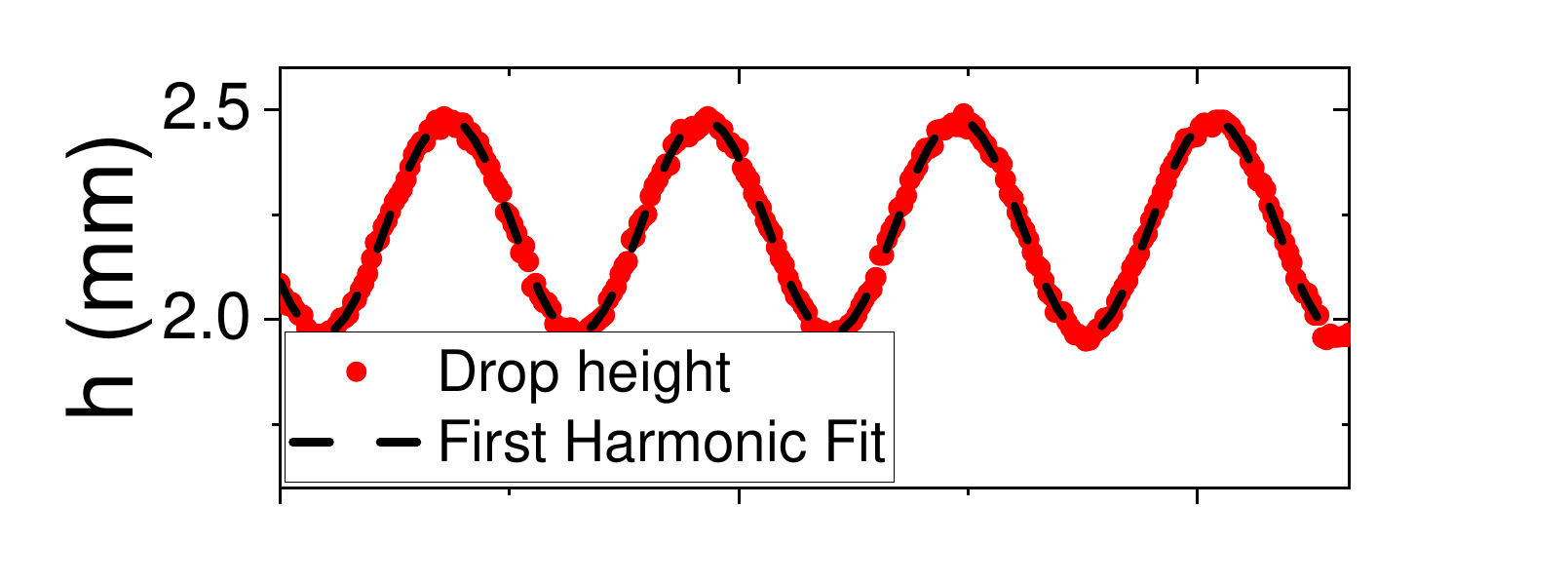}
    
    \includegraphics[width=0.85\linewidth]{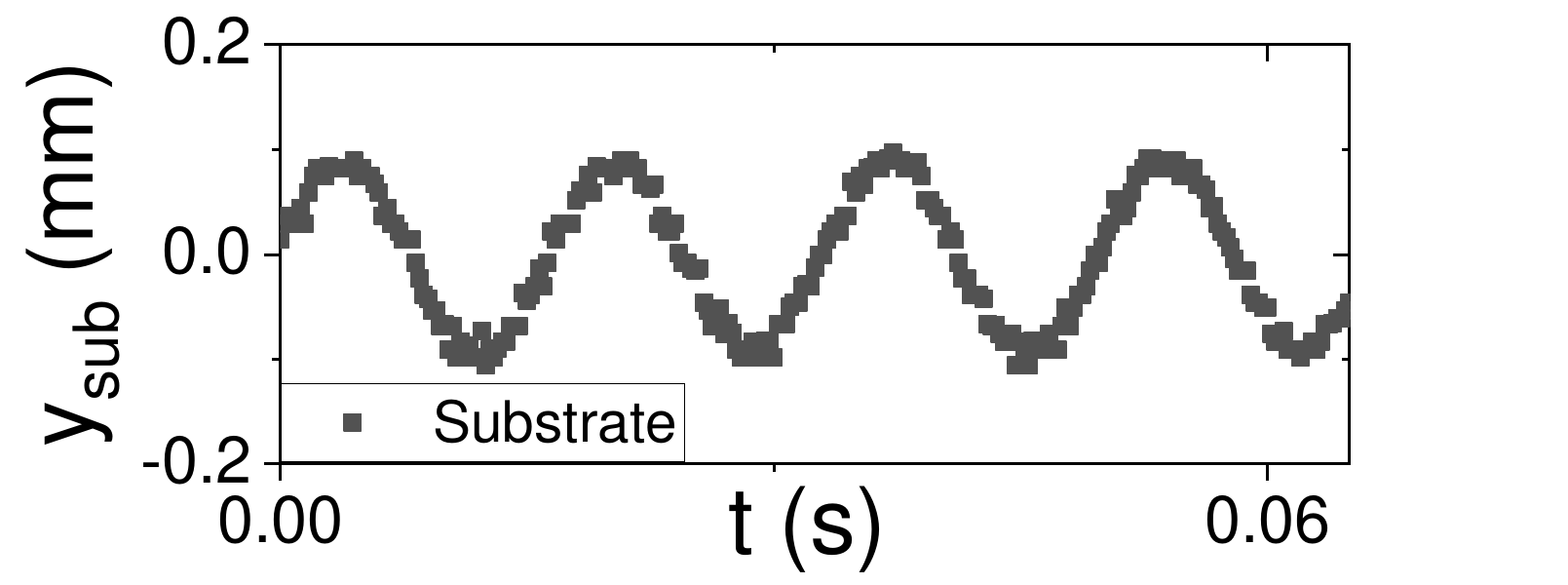}
    %\caption*{(A)}
  \end{subfigure}

  \caption{Top: Time evolution of the drop height and Bottom: the substrate position for a 3\% PEO solution drop (\SI{300}{\kilo\gram\per\mol}) subjected to oscillation at \SI{60}{\hertz}.
}
  \label{fig:PEO-300k}
\end{figure}

For example, when the substrate oscillates at \SI{60}{\hertz} and a PEO solution drop (3\%, \SI{300}{\kilo \gram \per \mol}) is placed on it, the drop height also oscillates at \SI{60}{\hertz} ($A=$\SI{0.1}{\milli \meter}).
The results, shown in Fig.~\ref{fig:PEO-300k}, display a clear phase lag ($\delta \approx$ \SI{150}{\degree}) and amplitude difference between the substrate motion and the drop response.
Fitting the drop height oscillation at this driving frequency yields the following expression:

\begin{equation}
h = 2.2 +0.25 \sin( 2\: \pi \:  60 \: t- 0.83\pi) 
 \label{LVE-response}
\end{equation}

\begin{figure}[htbp]
\centering
\begin{tikzpicture}

% Insert the images
\node at (0,6) {\includegraphics[width=5cm]{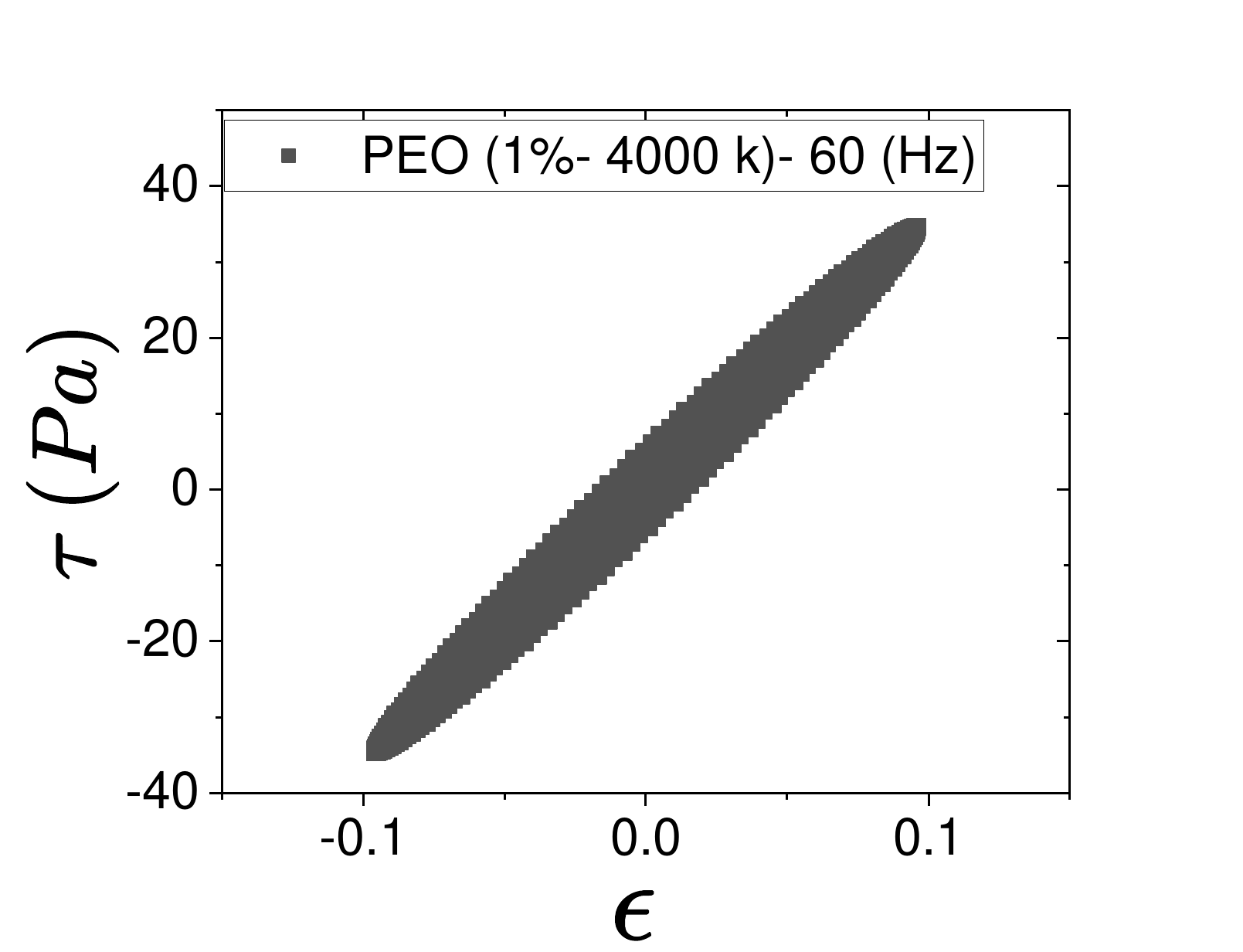}};
\node at (0,2.5) {\includegraphics[width=5cm]{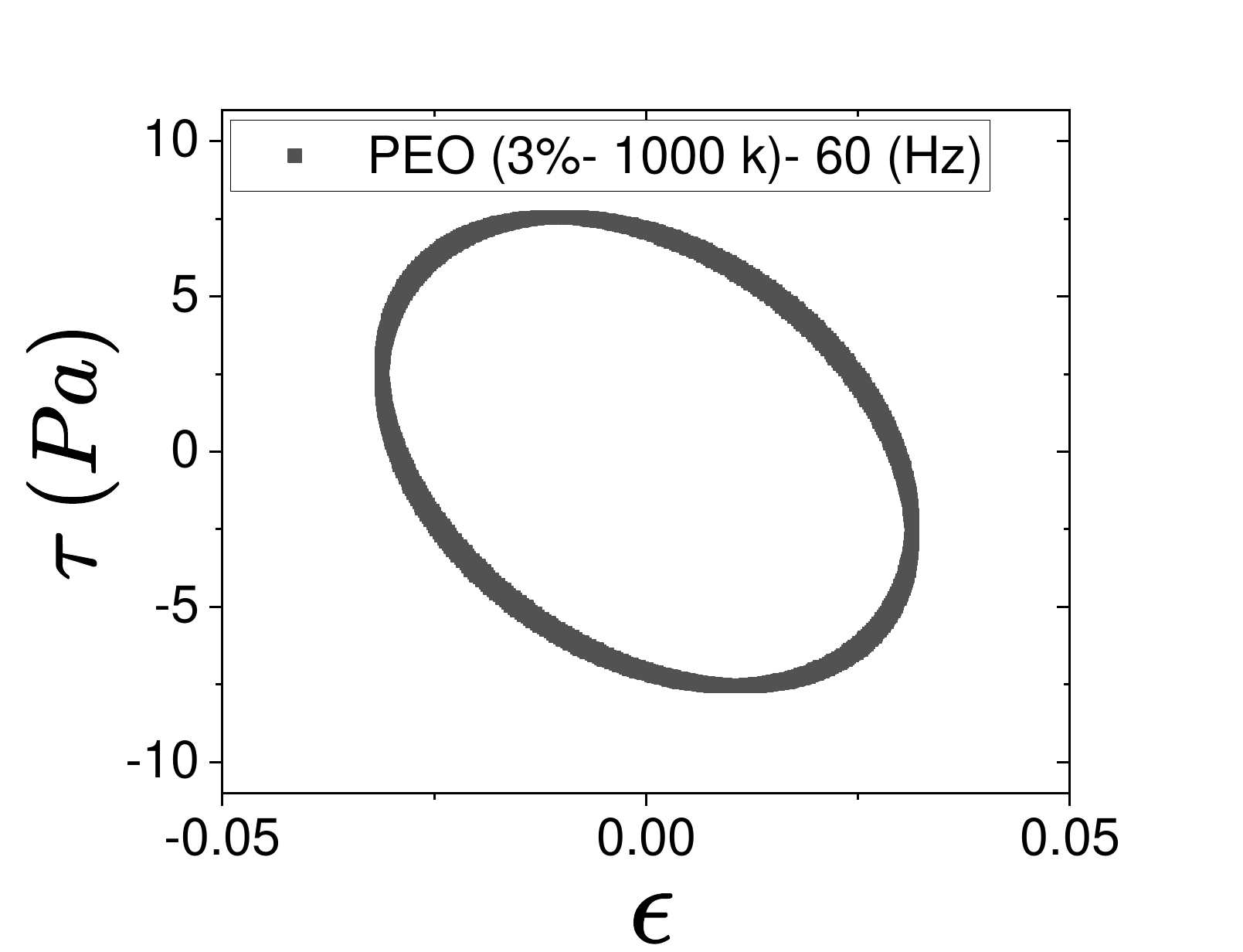}};
\node at (0,-1) {\includegraphics[width=5cm]{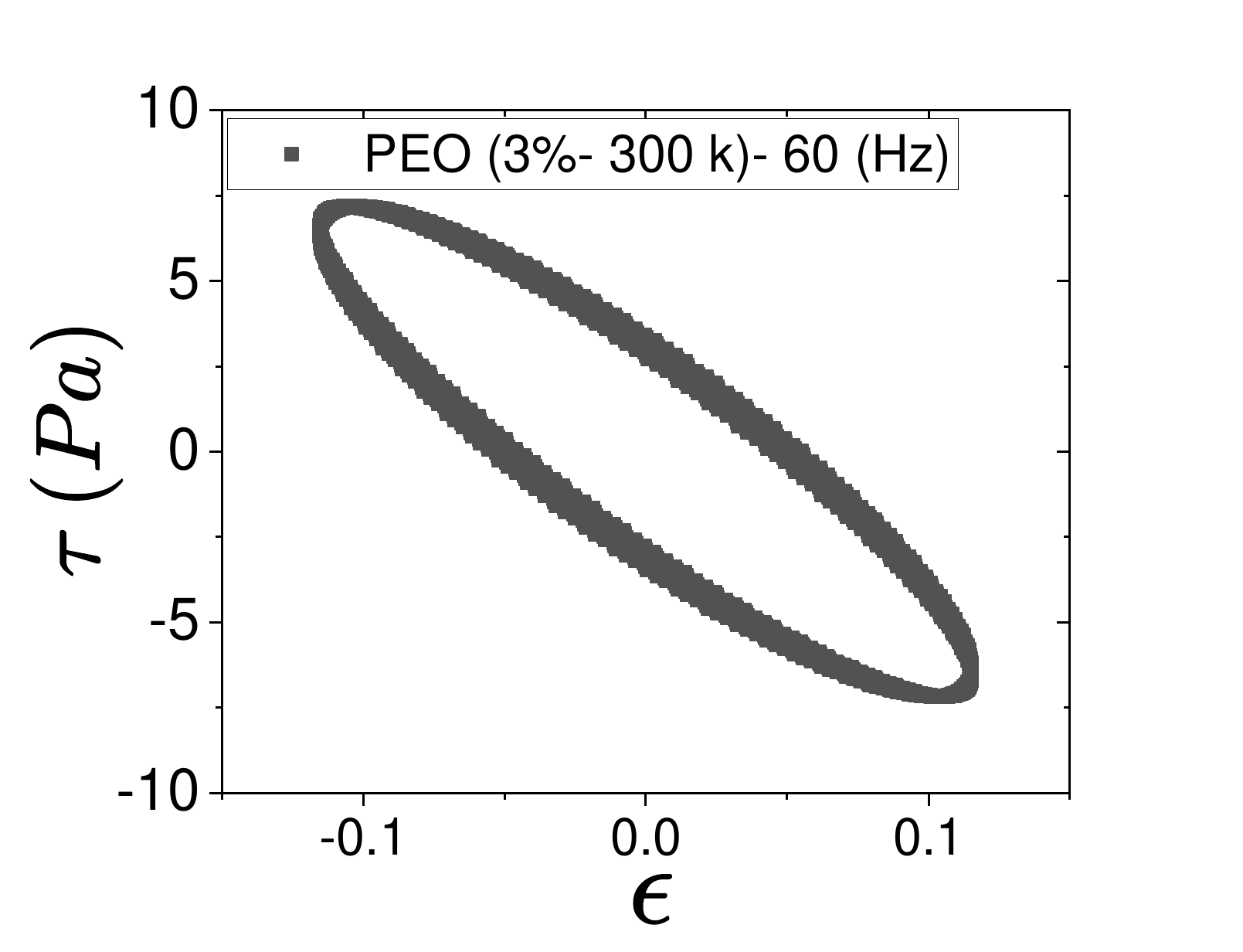}};

% Vertical arrow and labels
\draw[->, thick] (3,-2.5) -- (3,7.5);
\node at (2.5,-1) {0.01};
\node at (2.5,2.5) {5.5};
\node at (2.5,6) {70};

% Axis label
\node[rotate=90] at (3.7,3.25) { $\mathbf{De}$ };

\end{tikzpicture}
\caption{The Lissajous curves (Strain, $\epsilon$ - Stress, $\tau$) for different Deborah numbers (3\%, \SI{300}{\kilo\gram\per\mol}, 3\%, \SI{1000}{\kilo\gram\per\mol} and 1\%, \SI{4000}{\kilo\gram\per\mol}). The driving frequency for all three samples are \SI{60}{\hertz} and amplitude is $\approx$ \SI{0.5}{\milli \meter}. }
\label{De-Lissajou}
\end{figure}
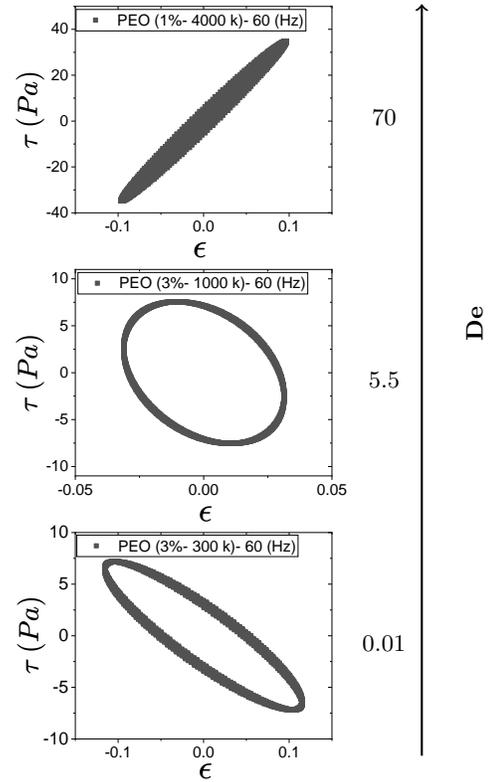

For the same drop experiment, we plot the Lissajous curves, by plotting strain versus stress (Fig.~\ref{De-Lissajou}, bottom).
The strain is assumed to be $\epsilon = \frac{h-h_{0}}{h_{0}}$.  
To smooth the data, we used the fitted sinusoidal function for the drop response.
The applied inertial stress due to substrate oscillation converts and dissipates through several sources:

\begin{itemize}
    \item Viscous dissipation.
    \item Capillary stress.
    \item Contact line dissipation.
    \item Elastic stress.
    \end{itemize}
    
The inertial stress is estimated by using Eq.~\ref{Inertia-Stress}.
For simplicity and because the detailed analysis of the remaining stress components is beyond the scope of this study, we have neglected them.

\begin{equation}
 \tau_{inertia} =\tau=-\rho A h \omega^2 sin(\omega t) 
 \label{Inertia-Stress}
\end{equation}

The physical interpretation of the Lissajous curves is based on the phase relationship between strain ($\epsilon=\epsilon_{0}\sin(\omega t)$) and stress ($\tau=\tau_{0}\sin(\omega t + \delta)$).
When strain and stress are in phase ($\delta = $ \SI{0}{\degree}), the curve forms a straight line, indicating a purely elastic response with no energy dissipation.
In contrast, when they are perfectly out of phase ($\delta = $ \SI{90}{\degree}), the curve forms a circle, corresponding to a purely viscous response.
For viscoelastic liquids in the linear regime, the response lies between these two extremes, resulting in an elliptical Lissajous curve ($0^{\circ}< \delta < 90^{\circ}$).
In drop experiments, both viscoelasticity and the damped driven oscillator contribute to the phase lag, which in turn influences the Lissajous curves.  
This behavior is observed in our experiments on the PEO solution, as shown in Fig.~\ref{De-Lissajou}, bottom.
This type of Lissajous curve is typically obtained through oscillatory shear measurements using commercial rheometers.
However, a known limitation of conventional rheometers is that, at high frequencies, the measurements become less reliable due to increasing inertial effects in the device.
In contrast, the drop experiments presented here allow for higher frequencies, making it particularly advantageous for exploring regimes that are relevant to many practical applications.

%From these curves (the time dependent strain and stress responses), we can calculate the  $G'$ (storage modulus) and $G''$ (loss modulus). 
%To do so, we have to consider $\epsilon=\epsilon_{0}\sin(\omega t)$ and $\tau=\tau_{0}\sin(\omega t+\phi)$, where $\epsilon_{0}$, $\tau_{0}$ and $\phi$ are the maximum strain, stress and phase lag.
%Then by having the phase lag ($\phi$), we can calculate the $G'$ and $G"$ as the following:

%\begin{equation}
%\begin{aligned}
 % G' &=  \frac{\tau_{0}}{\epsilon_{0}}\cos(\phi) \\
  %G" &= \frac{\tau_{0}}{\epsilon_{0}}\sin(\phi)
%\end{aligned}
%\end{equation}

To compare different viscoelastic liquids, we conducted experiments using samples with different rheological properties obtained by using different polymer concentrations and molar masses.
This allows us to investigate different levels of viscoelasticity.
To capture the effects of viscoelasticity, it is essential to couple the internal (material) and external (driving) dynamics.
For this purpose, we use the Deborah number ($De$), defined as the ratio of the polymer relaxation time to the characteristic external (experimental) time scale.
This dimensionless number was first introduced by Marcus Reiner \cite{reiner1964deborah, tschoegl2012phenomenological}, and the same concept was later used to define the elastocapillary number \cite{rostami2024spreading}.
In the present study, the Deborah number is defined as the ratio of the polymer relaxation time ($t_p$) to the inverse of the driving frequency, as shown in Eq.\ref{De-Number}.

\begin{equation}
 De =\frac{t_{p}}{1/f}=t_{p}f 
 \label{De-Number}
\end{equation}

Fig. \ref{De-Lissajou} presents Lissajous curves for three different cases: 3\% PEO (\SI{300}{\kilo\gram\per\mol}, $De=0.013$), 3\% PEO (\SI{1000}{\kilo\gram\per\mol}, $De=5.65$), and 1\% PEO (\SI{4000}{\kilo\gram\per\mol}, $De=69.6$).
The experiments are done at \SI{60}{\hertz}.
At low Deborah numbers ($De \ll 1$), the Lissajous curves exhibit classical viscoelastic behavior.
As the Deborah number increases to intermediate values ($De \approx 1$), the curves become more circular, indicating increased viscous dissipation.
The main reason behind this behavior is that around  $De \approx 1$, the polymer relaxation time matches the external driving time scale and the polymer can contribute to the viscosity \cite{ferry1980viscoelastic}.
In the high-$De$ regime ($De \gg  1$), strain and stress become more in phase ($\delta \to$ \SI{0}{\degree}), suggesting a nearly purely elastic response. 
This elastic behavior is consistent with the fact that the polymer relaxation time is much longer than the intrinsic time scale of the experiment, preventing the polymer chains from fully relaxing within each oscillation cycle.
In a completely different system, we have observed a similar dependence of drop dynamics on the matching of internal and external time scales in our previous studies on drop coalescence and spreading of polymer solutions \cite{rostami2024spreading, rostami2025coalescence}.

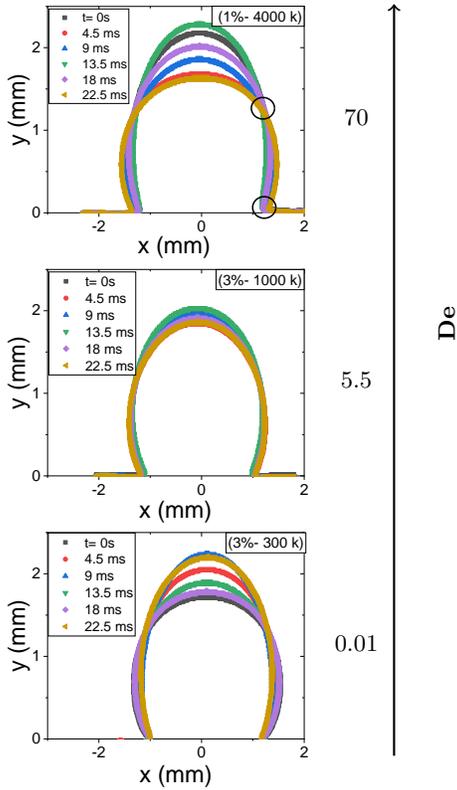
\begin{figure}[htbp]
\centering
\begin{tikzpicture}

% Insert the images
\node at (0,6) {\includegraphics[width=5cm]{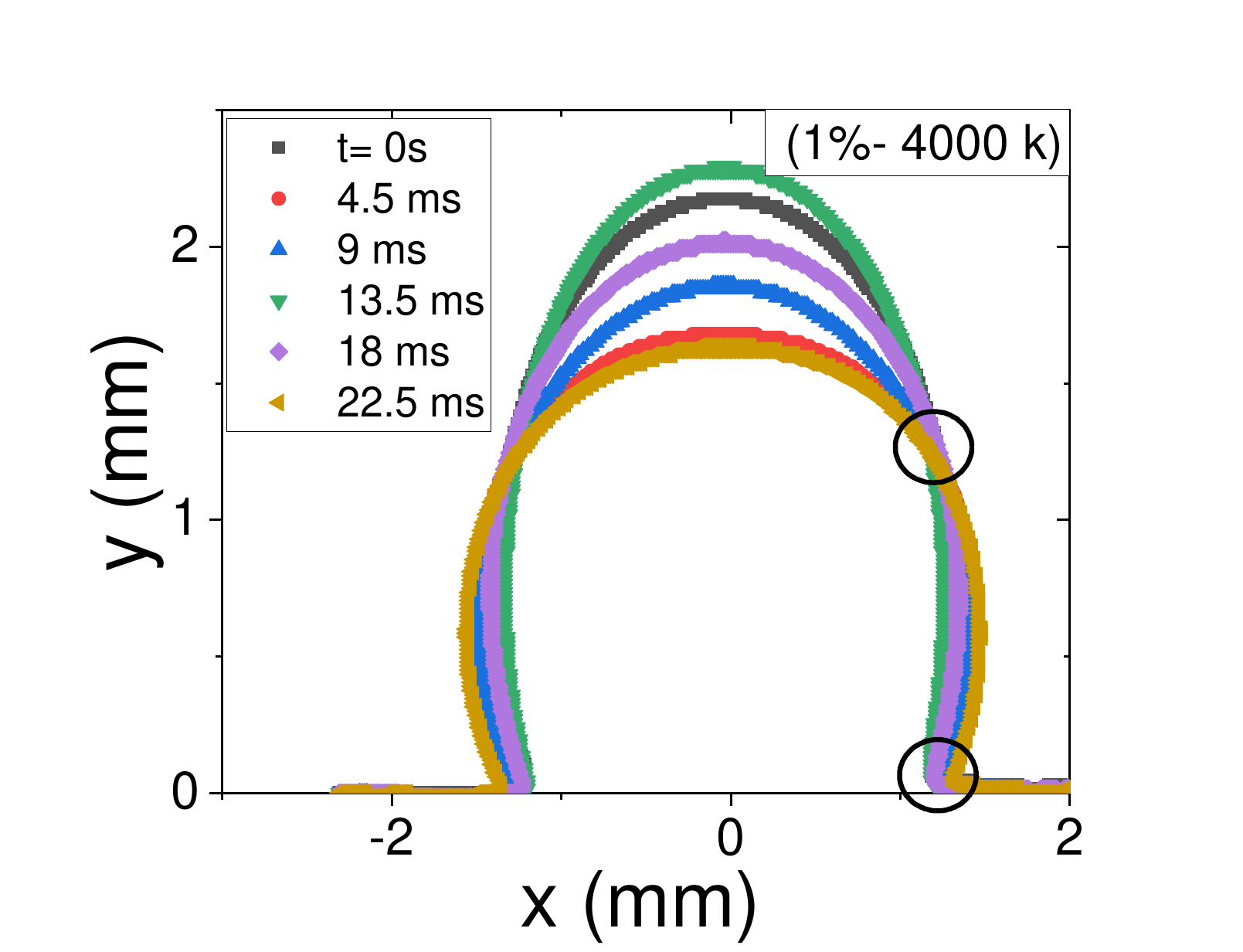}};
\node at (0,2.5) {\includegraphics[width=5cm]{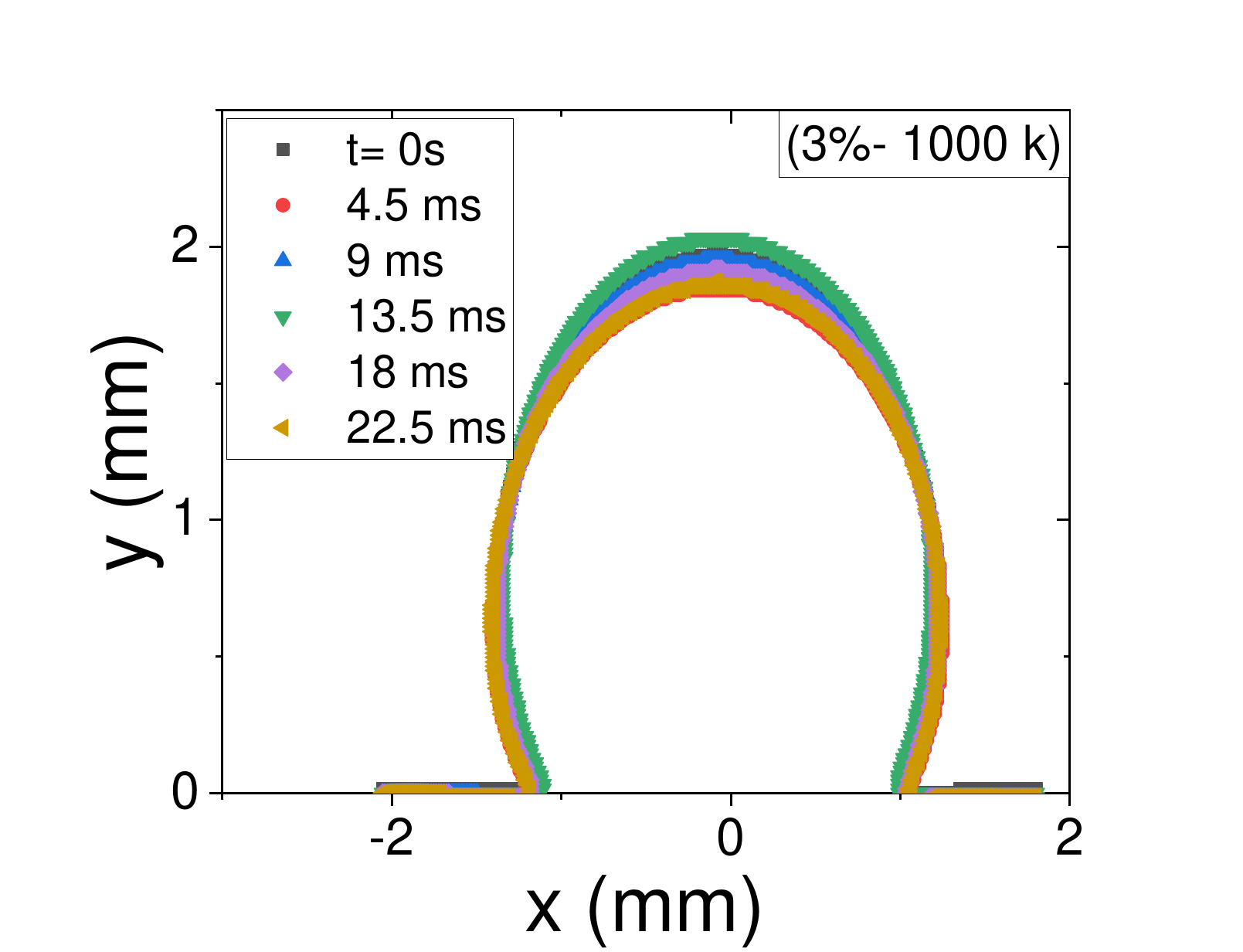}};
\node at (0,-1) {\includegraphics[width=5cm]{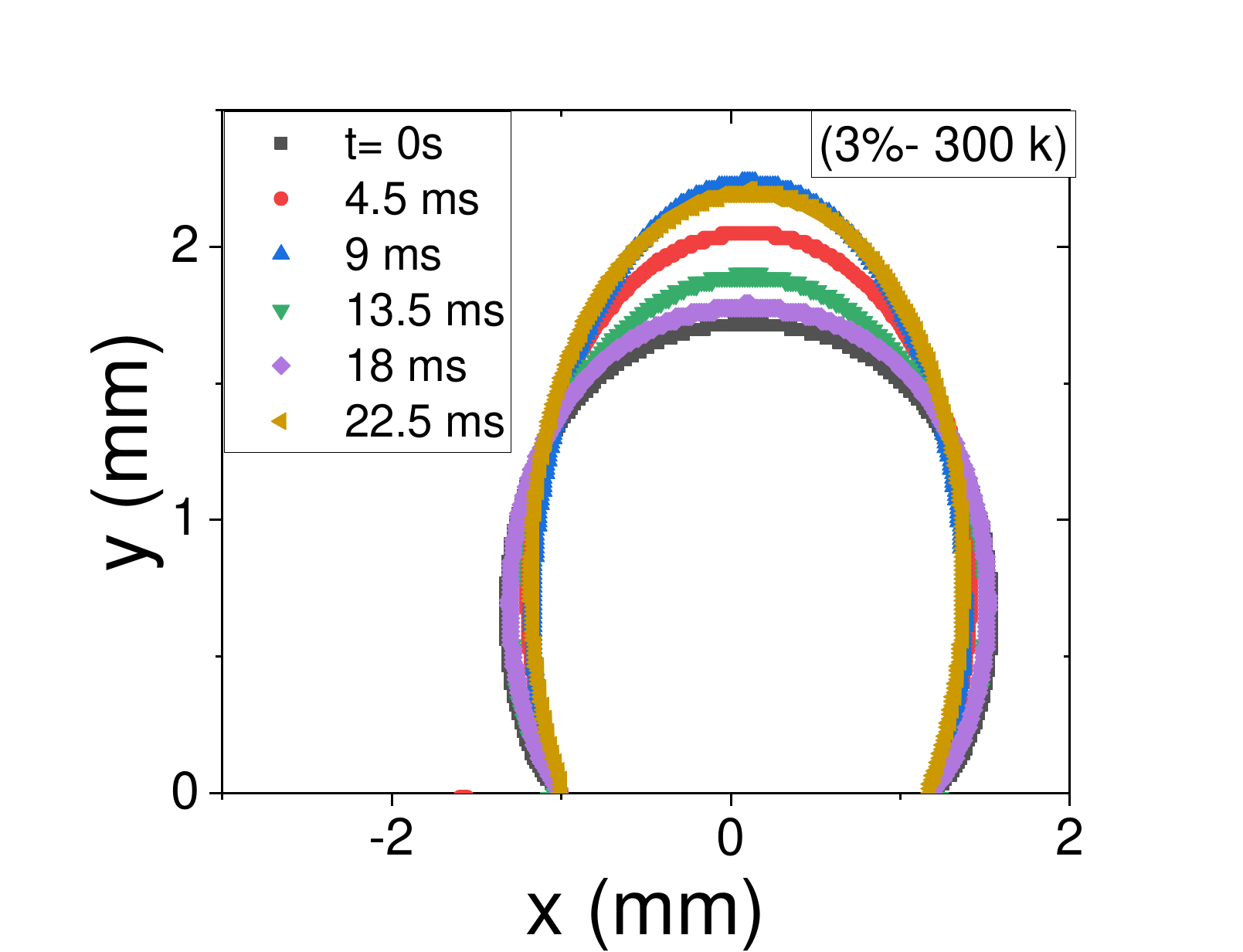}};

% Vertical arrow and labels
\draw[->, thick] (3,-2.5) -- (3,7.5);
\node at (2.5,-1) {0.01};
\node at (2.5,2.5) {5.5};
\node at (2.5,6) {70};

% Axis label
\node[rotate=90] at (3.7,3.25) { $\mathbf{De}$ };

\end{tikzpicture}
\caption{The drop profiles (interface) for different Deborah numbers (3\%, \SI{300}{\kilo\gram\per\mol}, 3\%, \SI{1000}{\kilo\gram\per\mol} and 1\%, \SI{4000}{\kilo\gram\per\mol}). The driving frequency for all three samples are \SI{60}{\hertz} and amplitude is $\approx$ \SI{0.5}{\milli \meter}.
In the top figure, two nodes (four by symmetry) are highlighted with circles, indicating the oscillation mode of $n = 2$. }
\label{De-Edge}
\end{figure}

Another indicator of drop dynamics is the oscillation mode, which can be extracted from the shape of the drop’s profile (i.e., the drop’s interface).
This method is based on spherical harmonics \cite{macrobert1967spherical} and has been previously applied to low-viscosity drops (e.g., water) by other groups \cite{steen2019droplet,noblin2004vibrated,chang2015dynamics,morozov2018vibration}.
For small-amplitude oscillations (i.e., small deformations), the dominant mode is the second mode ($n = 2$).
This indicates that, during oscillations, there are four nodes along the drop’s profile.
Two of these nodes are located at the pinned contact lines, while the other two occur at the intersections of successive drop profiles, as shown in Fig.~\ref{De-Edge}.
In all of our experiments up to now, both with linear viscoelastic regime and Newtonian drops, the oscillation mode remains constant at $n = 2$.
Although the mode number is unchanged, the drop outline and the amplitude of deformation vary depending on the drop properties.
As shown in Fig.~\ref{De-Edge}, the deformation amplitude is larger at both low and high Deborah numbers.
At high $De$ numbers, elastic effects dominate the response, while at very low $De$, the drop behaves similarly to a low-viscosity Newtonian liquid ($\zeta < 1$).
In contrast, for the intermediate case (3\%, \SI{1000}{\kilo\gram\per\mol}), the oscillation is strongly damped due to viscous dominance, consistent with the behavior observed in the corresponding Lissajous curves.

As mentioned earlier, one of the key advantages of this method is its ability to access broader range of frequencies.
To study the effect of frequency, we conducted experiments at various driving frequencies.
In general, for viscoelastic samples, increasing the frequency leads to flatter Lissajous curves, indicating a more elastic response \cite{ewoldt2008new}.
At higher frequencies, the polymers have less time to relax.

\begin{figure*}[htbp]
  \centering

  \begin{subfigure}[b]{0.45\textwidth}
    \includegraphics[width=\linewidth]{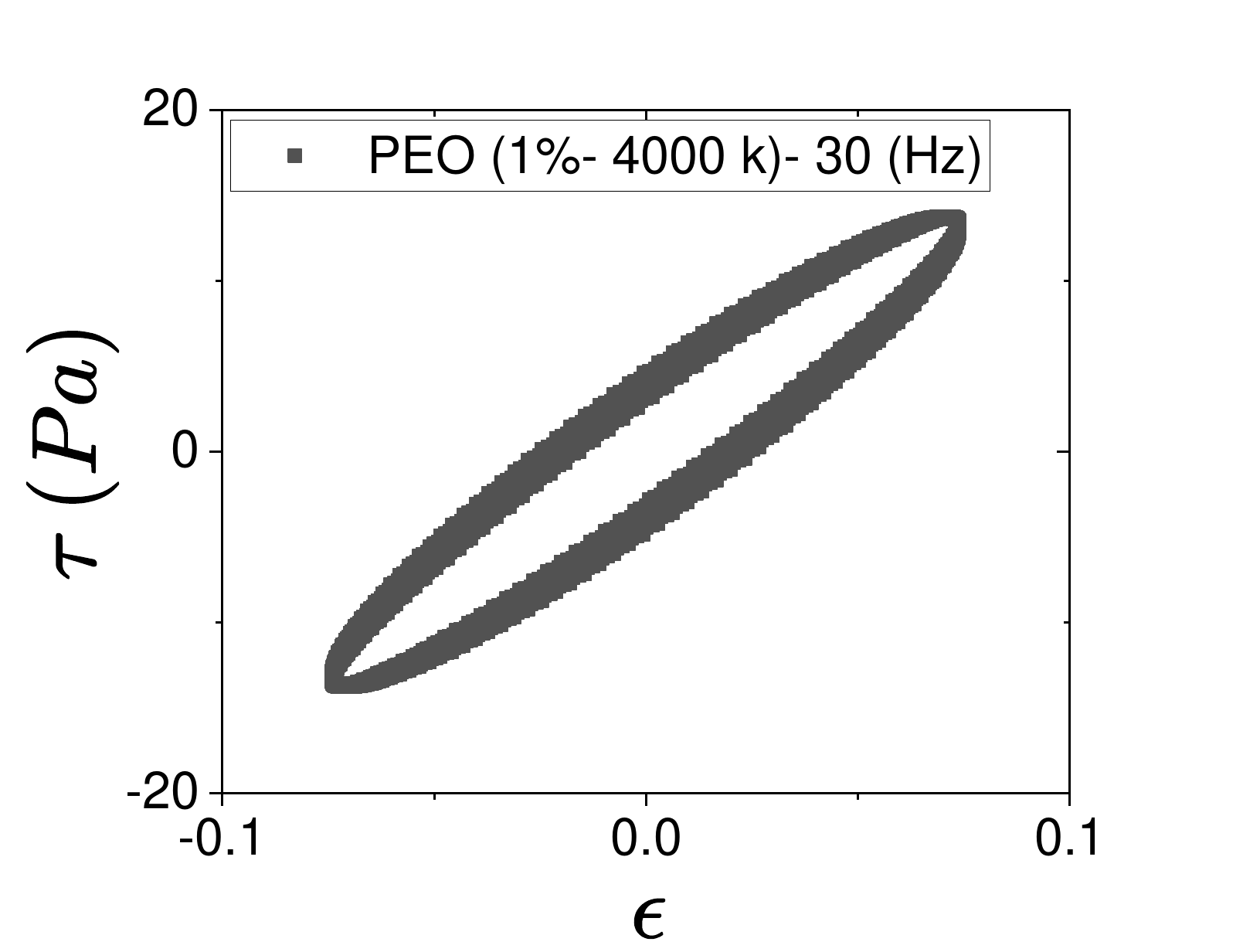}
    \caption{}
  \end{subfigure}
  \hfill
  \begin{subfigure}[b]{0.45\textwidth}
    \includegraphics[width=\linewidth]{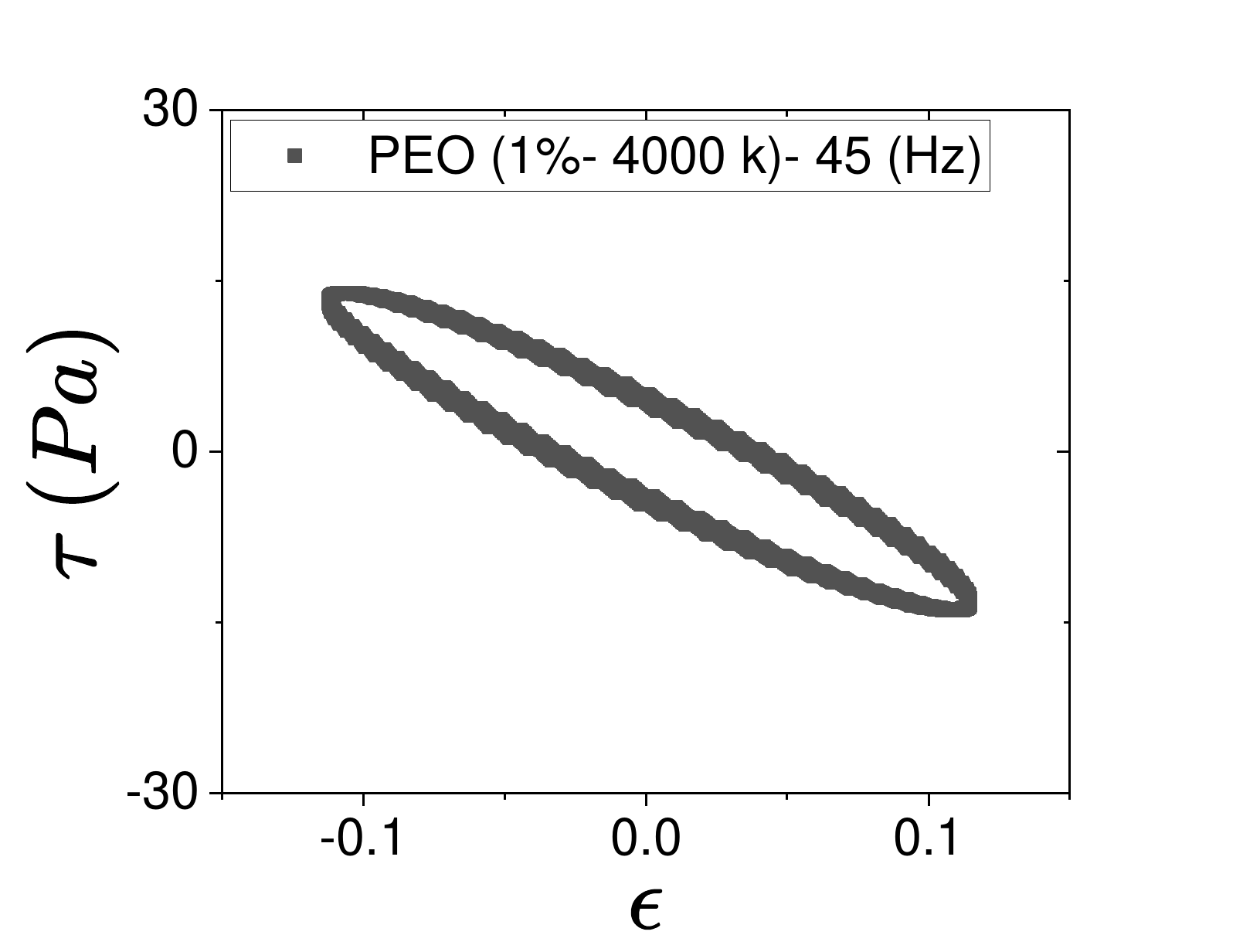}
    \caption{}
  \end{subfigure}
  
  \vspace{1em}

  \begin{subfigure}[b]{0.45\textwidth}
    \includegraphics[width=\linewidth]{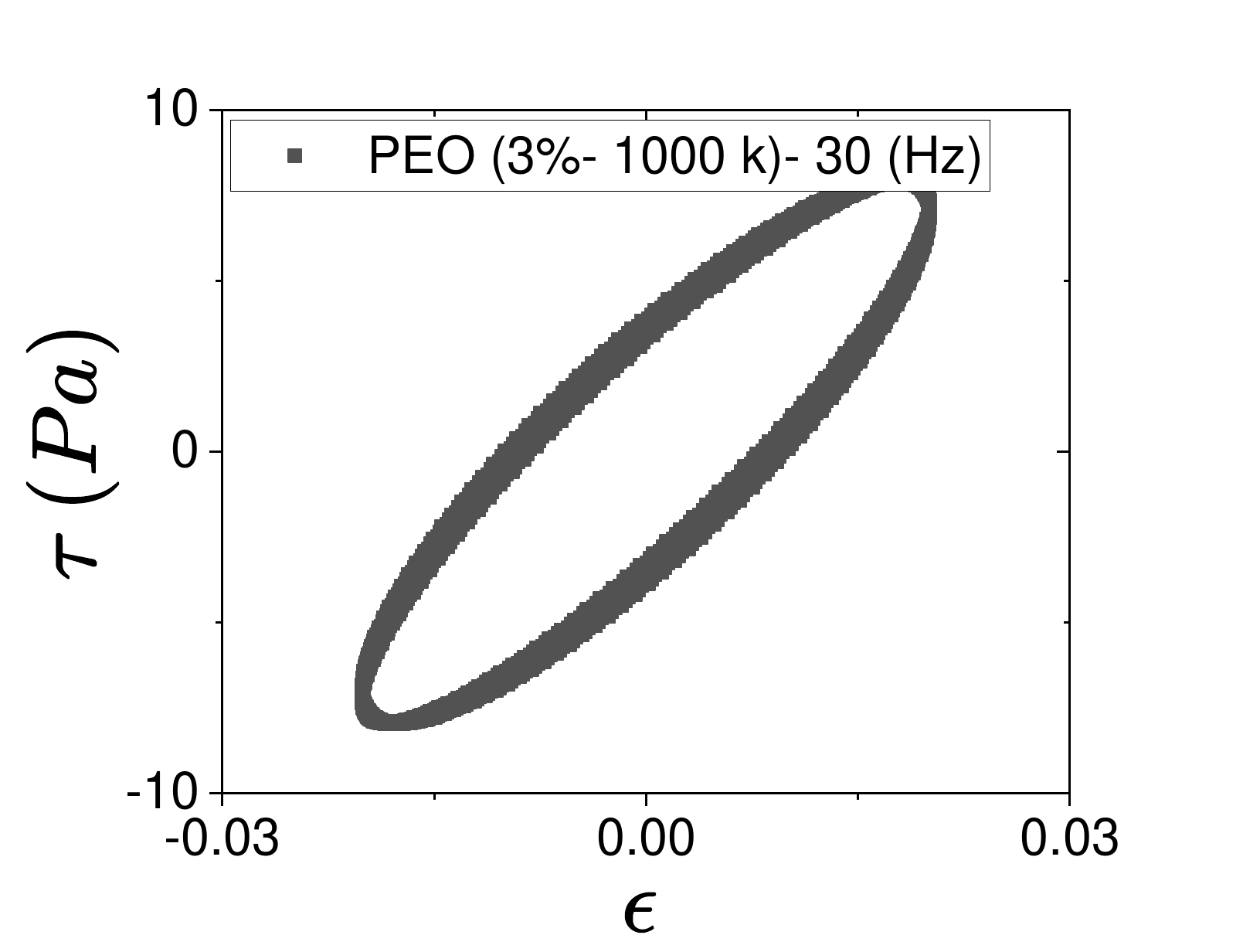}
    \caption{}
  \end{subfigure}
  \hfill
  \begin{subfigure}[b]{0.45\textwidth}
    \includegraphics[width=\linewidth]{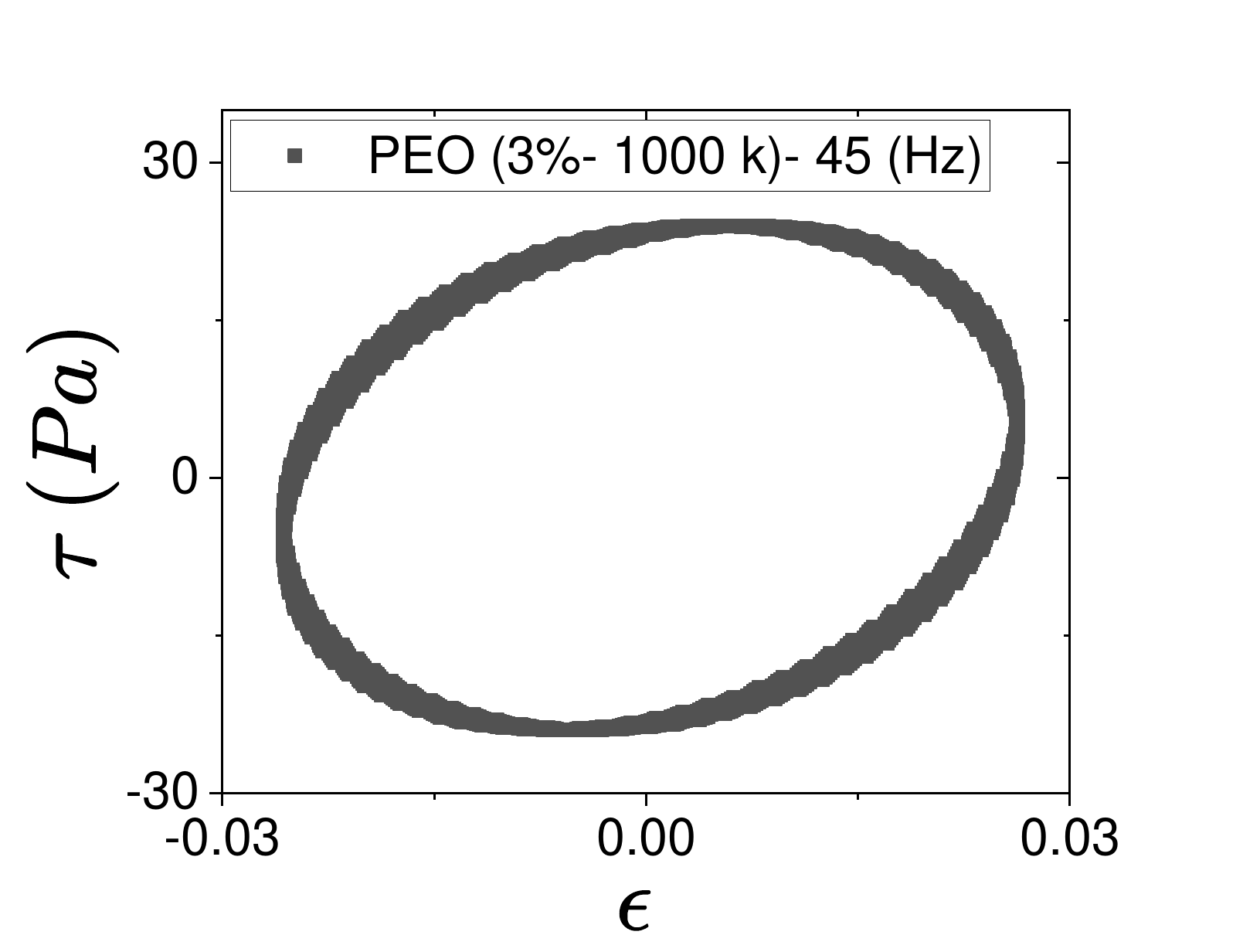}
    \caption{}
  \end{subfigure}

  \caption{The Lissajous curves (Strain, $\epsilon$ - Stress, $\tau$) for a) 1\%, \SI{4000}{\kilo\gram\per\mol} at \SI{30}{\hertz}, ($De=35$) b) 1\%, \SI{4000}{\kilo\gram\per\mol} at \SI{45}{\hertz}, ($De=52$) as well as c) 3\%, \SI{1000}{\kilo\gram\per\mol} at \SI{30}{\hertz}, ($De=2.8$) and d) 3\%, \SI{1000}{\kilo\gram\per\mol} at \SI{45}{\hertz}, ($De=4.2$).}
  \label{De-Frequency}
\end{figure*}

It should be noted that the eigenfrequency of the drop itself (obtained from experimental dynamics) also influences the drop response.
In general, there are two distinct contributions to the phase lag: the first arises from viscoelasticity, and the second originates from the damped driven oscillator model.
Decoupling these two effects lies beyond the scope of the present paper; however, one possible approach is to perform measurements at off-resonant frequencies and employ physics-informed, data-driven models to predict the behavior at the resonant frequency.
As the phase lag at the eigenfrequency is \SI{90}{\degree}, the dissipation is maximized, and the Lissajous curve becomes more circular.
To examine this behavior, we plotted Lissajous curves for two representative cases: a viscous-dominated sample (3\%, \SI{1000}{\kilo\gram\per\mol}) and an elastic-dominated sample (1\%, \SI{4000}{\kilo\gram\per\mol}).
Results for two driving frequencies, \SI{30}{\hertz} and \SI{45}{\hertz}, are shown in Fig.~\ref{De-Frequency}.
For the highly elastic case (1\%, \SI{4000}{\kilo\gram\per\mol}), the Lissajous curves become noticeably flatter when comparing the response at \SI{30}{\hertz} (Fig.\ref{De-Frequency}) to that at \SI{60}{\hertz} (Fig.\ref{De-Lissajou}), consistent with increased elastic behavior at higher frequencies.
In contrast, for the viscous-dominated sample (3\%, \SI{1000}{\kilo\gram\per\mol}), the shape of the Lissajous curves remains largely unchanged with increasing frequency, reflecting the viscous dominance of polymer solutions.

\begin{figure}[htbp]
\includegraphics[width=0.95\linewidth]{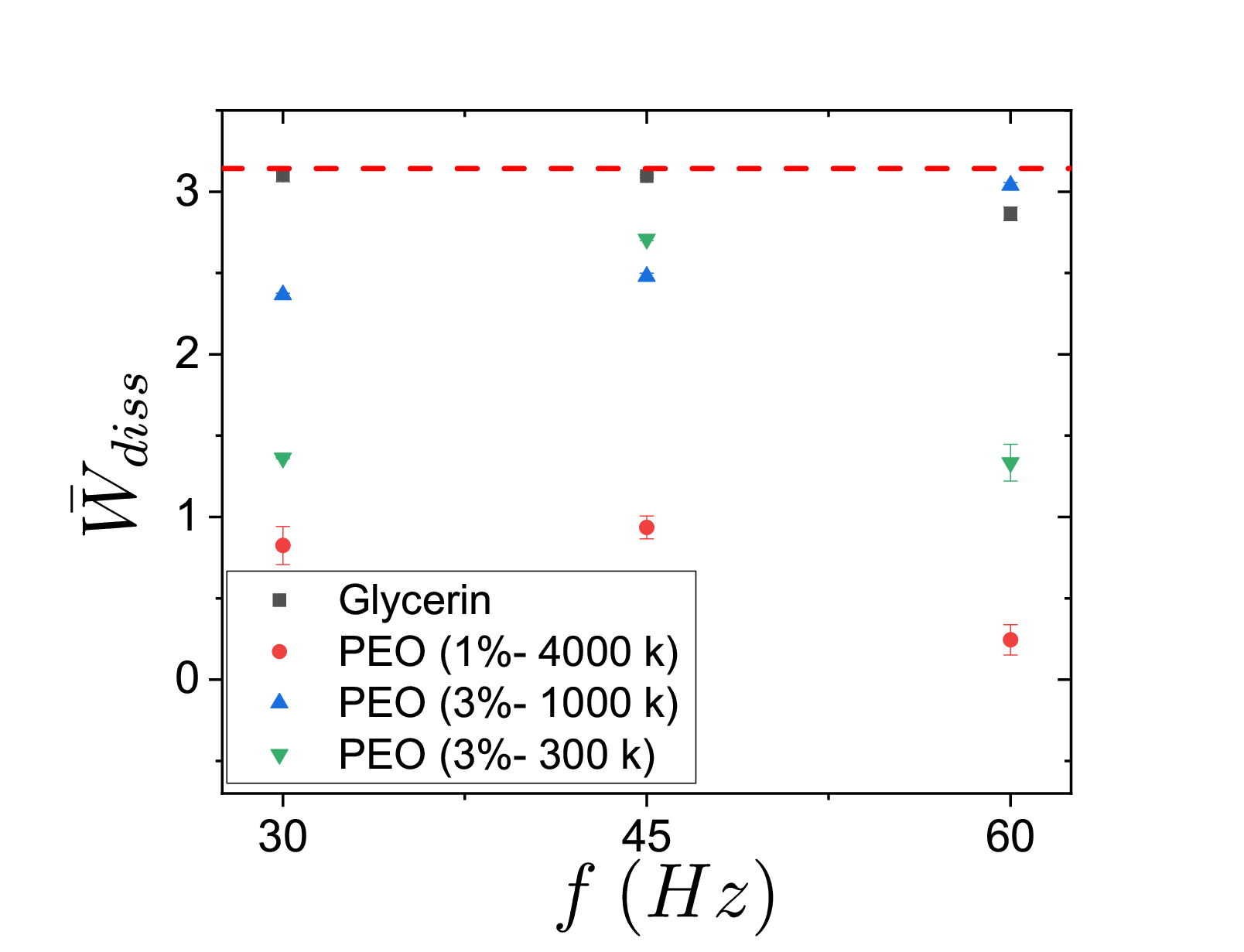}% Here is how to import EPS art
\caption{\label{Wdiss-Frequency} The normalized dissipated energy ($\bar{W}_{\text{diss}}$) for the samples listed in Table~\ref{Table-Properties}, measured at three driving frequencies: 30, 45 and \SI{60}{\hertz}.}
\end{figure}

The area enclosed by a Lissajous curve represents the energy dissipated during one deformation cycle.
This dissipated energy is calculated as $W_{\text{diss}} = \oint \tau d\epsilon$.
A larger enclosed area corresponds to greater energy dissipation.
To enable comparison across different experiments, both stress and strain were normalized by their respective maximum values ($\bar{\tau}=\tau/\tau_{\text{max}}$ and $\bar{\epsilon}=\epsilon/\epsilon_{\text{max}}$).
The normalized dissipated energy is then computed as ($\bar{W}_{\text{diss}}= \oint \bar{\tau} d\bar{\epsilon}$). 
Results for all samples and three driving frequencies are summarized in Fig.~\ref{Wdiss-Frequency}.

%\begin{figure}
%\includegraphics[width=0.85\linewidth]{De-1per-4M-3per-1M.png}% Here is how to import EPS art
%\caption{\label{De-Frequency} A sketch of drop oscillating experimental set-up, which has shaker and recording system.}
%\end{figure}

As mentioned earlier, each cycle of the stress–strain Lissajous curves for our liquids takes an elliptical form, with the maximum enclosed area corresponding to a circle. 
Since we use normalized data ($-1 < \tau < 1$ and $-1 < \epsilon < 1$), the maximum surface area is given by $\bar{W}{\text{diss}} = \pi r^2 = \pi$. 
In Fig.~\ref{Wdiss-Frequency}, this maximum value is shown as red dashed lines. 
As illustrated, the measured values for pure glycerin are very close to this maximum, indicating the dominance of viscous dissipation. 
Moreover, the closer the surface area approaches the maximum value ($\bar{W}{\text{diss}} \to \pi$), the more circular the Lissajous curve becomes. 
For the 3\%, \SI{300}{\kilo\gram\per\mol} drop, a peak in $\bar{W}_{\text{diss}}$ is observed around \SI{45}{\hertz}.
We attribute this peak to the critical frequency identified earlier for the water drop (see Fig.\ref{Water-Frequency-Sweep} and Fig.\ref{Strain-Frequency}).
This explanation is plausible, as the polymer in this sample is relatively short and at low concentration, making its behavior more similar to the Newtonian liquid ($\zeta < 1$ more dissipation at critical frequency).
For the 3\%, \SI{1000}{\kilo\gram\per\mol} drop, $\bar{W}_{\text{diss}}$ increases approximately linearly with frequency, which is consistent with the behavior expected from viscous-dominated fluids.
The absolute values of $\bar{W}_{\text{diss}}$ are close to the viscous case.
In contrast, for high molar mass samples—1\%, \SI{4000}{\kilo\gram\per\mol}—$\bar{W}_{\text{diss}}$ remains nearly constant and small, with a small peak near \SI{45}{\hertz}, reflecting the dominant elastic response of these fluids.

\subsection{Nonlinear viscoelastic regime }

\begin{figure}[htbp]  % or [htbp] or [H] if using float package
  \centering
  \includegraphics[width=\columnwidth]{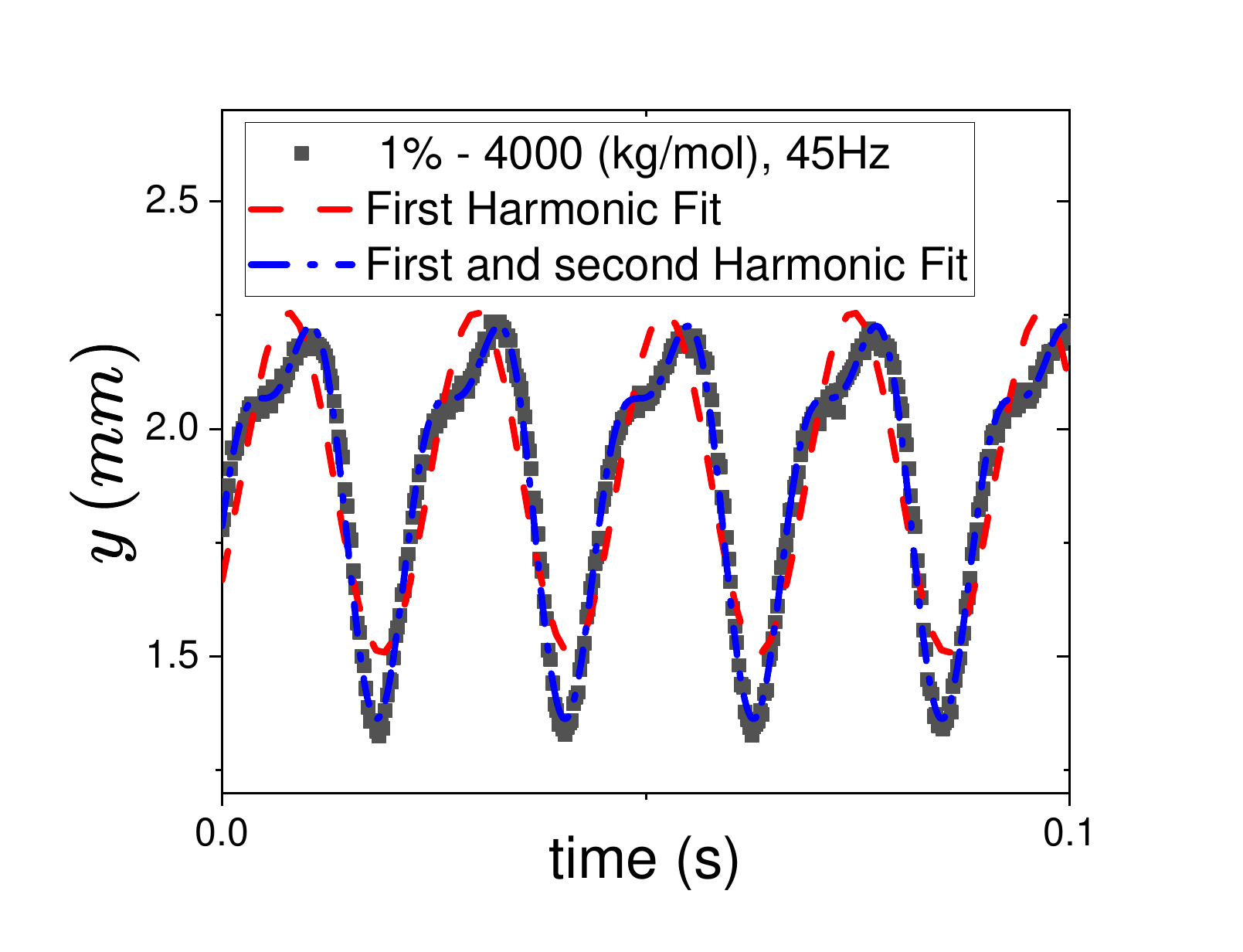} \\

  \includegraphics[width=\columnwidth]{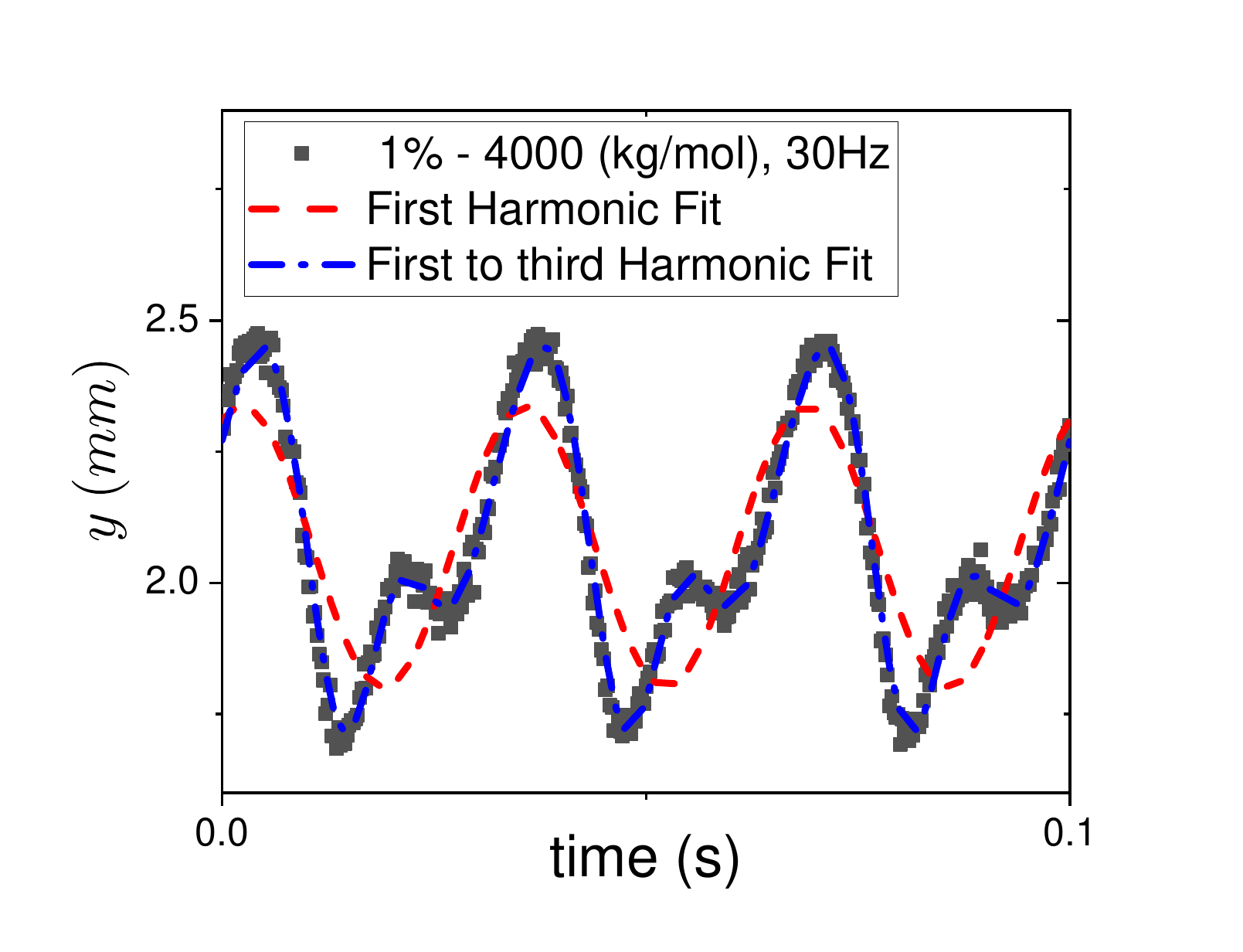}
  \caption{Time evolution of the drop height for a PEO solution drop (1\%, \SI{4000}{\kilo\gram\per\mol}) at two driving frequencies: top, \SI{45}{\hertz}; bottom, \SI{30}{\hertz}.
The red dashed line represents a fit using only the first harmonic ($\propto \sin(\omega t)$), while the blue dash-dotted line includes higher harmonics ($\propto \sum_n \sin(n\omega t)$). }
  \label{higher-harmonics}
\end{figure}

When the forcing amplitude is increased, the drop behavior becomes nonlinear, and the response exhibits higher harmonics, described by ($ h (t)=\sum_{n}^{} A_{n}\sin(n\omega t+\delta _{n})$).
To demonstrate this behavior, we analyzed the response of a 1\%, \SI{4000}{\kilo\gram\per\mol} drop at two driving frequencies: \SI{30}{\hertz} and \SI{45}{\hertz}.
Compared to the previous section, the drop deformation ($\epsilon$) was increased to 20\% by raising the driving amplitude ($A$).
As shown in Fig.~\ref{higher-harmonics}, the drop response clearly deviates from a simple sinusoidal form with a single dominant frequency ($\propto \sin(\omega t)$).
At \SI{45}{\hertz}, including the second harmonic term ($\sin(2\omega t)$) significantly improves the fit, yielding $R^2 = 0.992$.
At \SI{30}{\hertz}, incorporating the first, second, and third harmonics results in a fit quality of $R^2 = 0.987$.
The number of required harmonics depends on the driving frequency and amplitude, as well as on the viscoelasticity of the drop.
The presence of higher harmonics in the drop response provides clear evidence that the system has entered the nonlinear viscoelastic regime.

\begin{figure}[htbp] % or [htbp] or [H] if using float package
  \centering
  \includegraphics[width=\columnwidth]{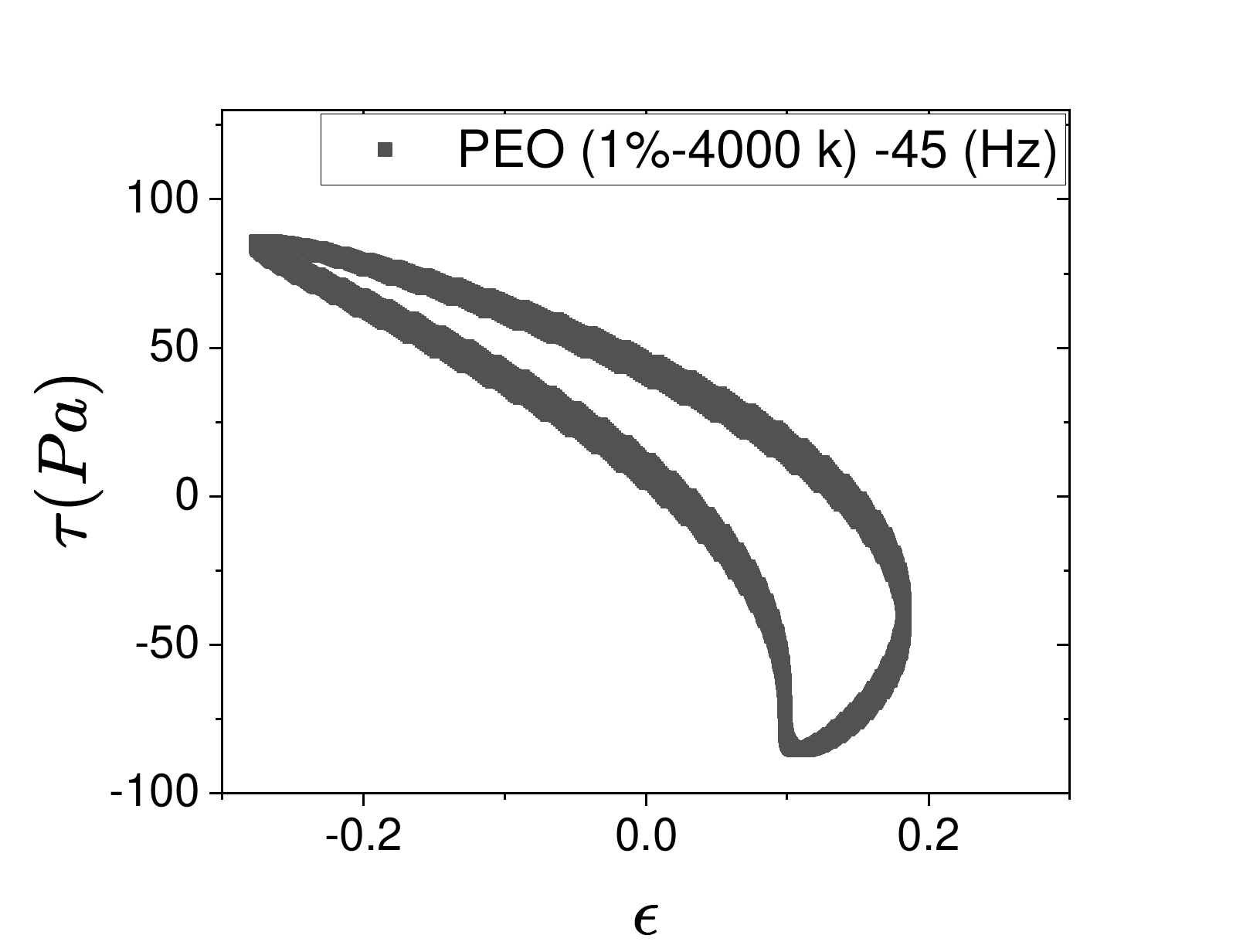} \\

  \includegraphics[width=\columnwidth]{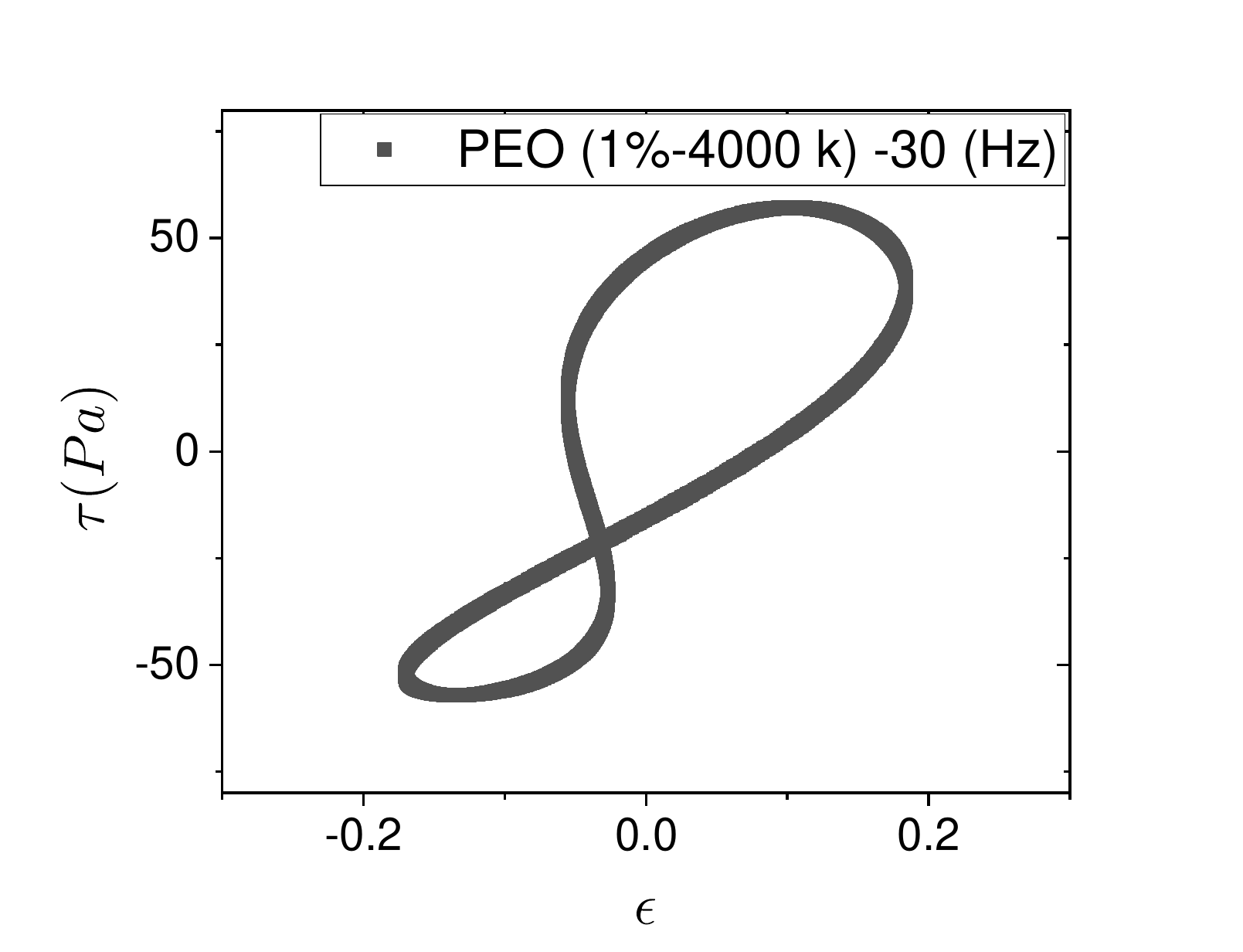}
  \caption{The Lissajous curves (Strain, $\epsilon$-Stress, $\tau$) for PEO 1\%, \SI{4000}{\kilo\gram\per\mol} at two driving frequencies: top, \SI{45}{\hertz}; bottom, \SI{30}{\hertz}.}
  \label{higher-harmonics-Lissajous}
\end{figure}

As mentioned in the introduction, in conventional rheology, even harmonics are typically absent due to the inherent symmetry of both the applied deformation and the material response ($\tau = \sum_{n=1,3,5,\dots} \tau_n \sin(\omega_{n} t + \delta_{n})$).
From this relation, we can extract the storage modulus ($G'$), representing the elastic component, and the loss modulus ($G''$), representing the viscous component.
However, in drop experiments, such symmetry does not necessarily hold, primarily due to geometric asymmetries and free surface effects.
Such asymmetry explains the presence of $even$ harmonics in the drop response.
The appearance of a second harmonic, in particular, leads to an asymmetric Lissajous curve, deviating from the symmetric elliptical shape observed in the linear regime, Fig.\ref{fig:PEO-300k}.
For the same experiments (Fig.\ref{higher-harmonics}), the corresponding Lissajous curves are shown in Fig.~\ref{higher-harmonics-Lissajous}.
At \SI{45}{\hertz}, the curve exhibits noticeable distortion, as expected in the nonlinear regime (second harmonic).
When third harmonics are present in the response, the Lissajous curve takes the form of a self-intersecting, figure-eight–like shape.
This behavior has also been reported in previous studies.
In our case, the cross point of eight-like curve is not at the center ($0,0$), which we attribute to the influence of the substrate.
Specifically, while the drop can oscillate freely away from the substrate, its motion is restricted when it is pushed toward the substrate, limiting its ability to deform symmetrically.

\begin{figure}
\includegraphics[width=0.85\linewidth]{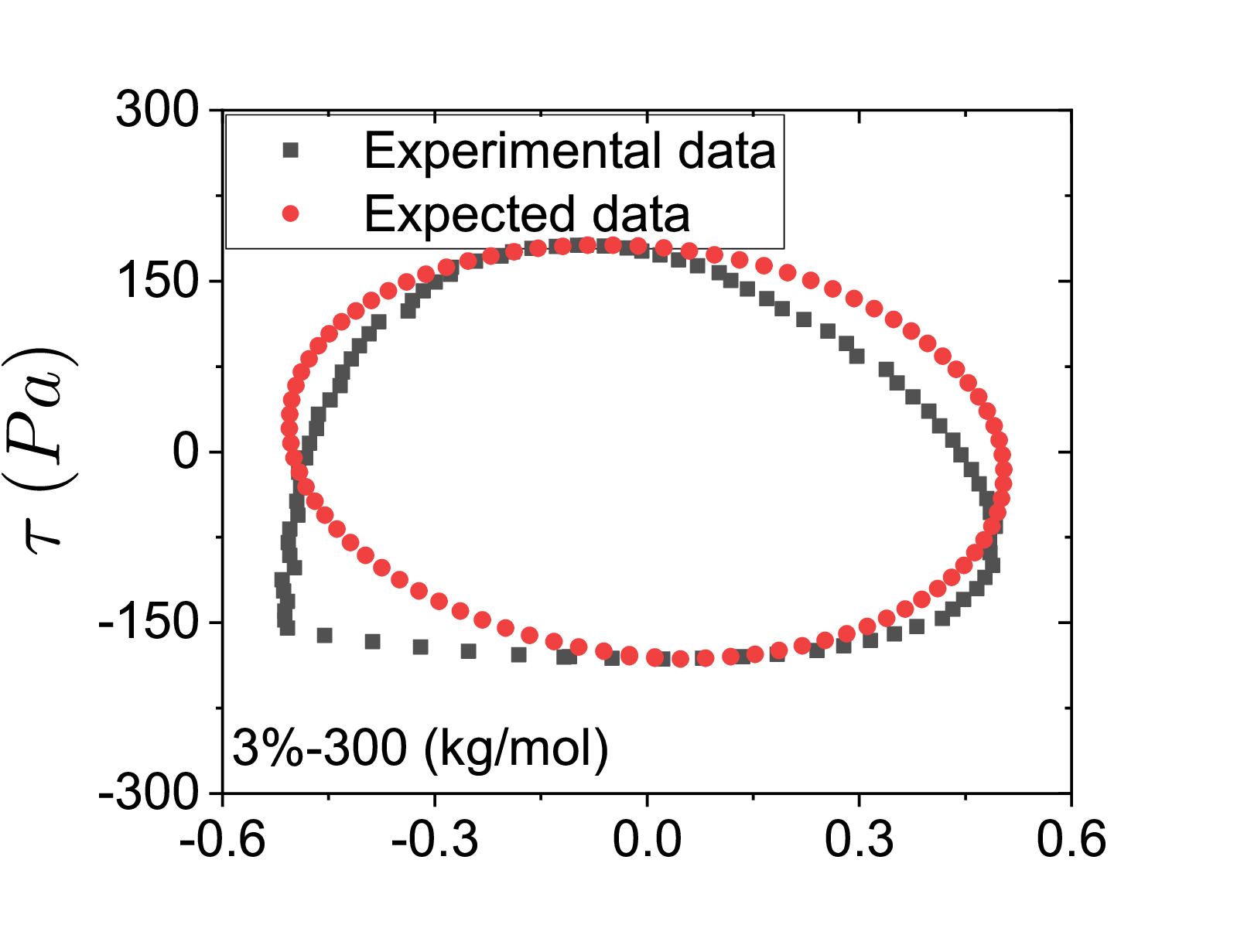}% Here is how to import EPS art
\caption{\label{Lissajous-3per-300k} The Lissajous curves (strain, $\epsilon$–stress, $\tau$) for PEO 3 \%, \SI{300}{\kilo\gram\per\mol}, at \SI{45}{\hertz}. Black squares represent the experimental data, while red circles correspond to the first harmonic fit $\sin(\omega t)$. The deviation between the experimental and fitted data arises from the presence of the substrate and the fact that the drop cannot penetrate the solid surface.}
\end{figure}

To only study the effect of the substrate, a drop of PEO (3 \%, \SI{300}{\kilo\gram\per\mol}) is subjected to vertical vibration with an amplitude of \SI{1.2}{\milli\meter} at \SI{45}{\hertz}.
In this case, the drop response should be linear, since the Deborah number is smaller than unity and the elastic contribution is negligible.
The corresponding Lissajous curve is shown in Fig.~\ref{Lissajous-3per-300k}. 
The asymmetric deformation of the drop in the negative deformation region leads to a distortion of the Lissajous curve.
For high Deborah numbers ($De > 1$) and large-amplitude deformations ($\epsilon > 30\%$), when the nonlinear viscoelastic regime and the substrate effect occur simultaneously, the Lissajous curves become strongly distorted.

\begin{figure}
\includegraphics[width=0.85\linewidth]{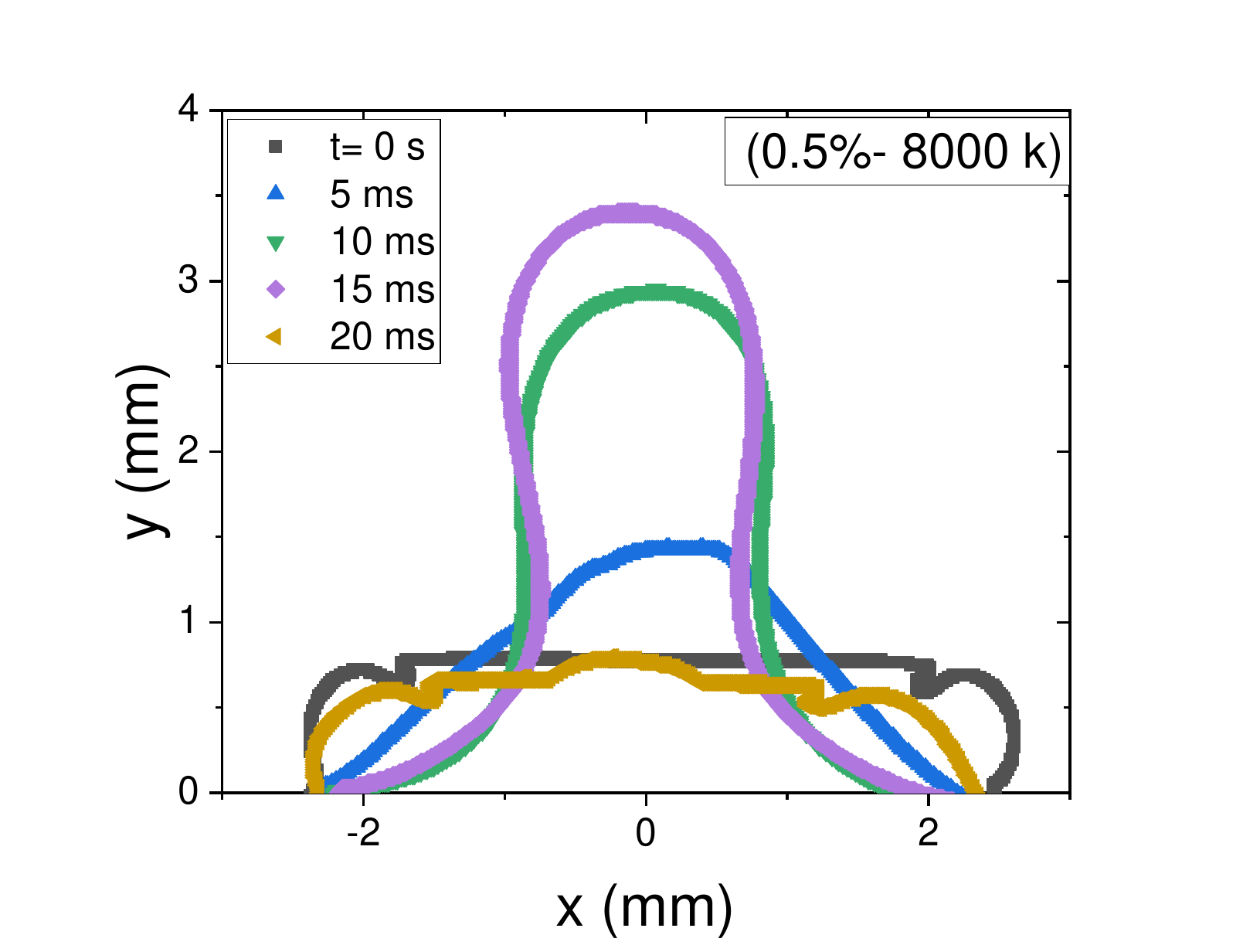}% Here is how to import EPS art
\caption{\label{05per-8M-dropshape} The drop profiles (interface) for 0.5\%, \SI{8000}{\kilo\gram\per\mol}. 
The driving frequency is \SI{45}{\hertz} and amplitude is $\approx$ \SI{1.3}{\milli \meter}.
The time evolution from 0 to \SI{20}{\milli \second} is presented.
The straight line in the black line is due to the dipole of the drop. 
The apparent stair-step features along the black and golden lines are artifacts arising from the image analysis .}
\end{figure}

The final point concerns elastic drop deformations under large-amplitude oscillations.
Figure~\ref{05per-8M-dropshape} shows the time evolution of the drop interface for a 0.5% PEO sample (\SI{8000}{\kilo\gram\per\mol}) oscillating at \SI{45}{\hertz} with an amplitude of \SI{1.3}{\milli\meter}.
The first notable observation is that the drop interface becomes asymmetric over successive oscillation cycles.
Second, the oscillation mode is no longer dominated by the second harmonic ($n \neq 2$), indicating a transition to more complex deformation modes.
Lastly, when the drop is pushed toward the substrate, the influence of the solid boundary becomes more pronounced, as the drop cannot oscillate freely in the negative $y$-direction.
In these images, the presence of a central dimple prevents complete characterization of the deformation using a single camera (see Fig.~\ref{05per-8M-dropshape}).
This confinement effect further contributes to the asymmetry and nonlinear response.
A detailed investigation of the nonlinear viscoelastic regime and the corresponding Lissajous curves lies beyond the scope of the present work and will be addressed in future studies.

\section{\label{sec:Results} Conclusion:}

We studied the oscillatory behavior of liquid drops placed on a vertically vibrating hydrophobic substrate, using side-view high-speed imaging to capture the drop response.
A wide range of polymer solutions—with varying concentrations and molecular weights—as well as Newtonian fluids (water and pure glycerin) were investigated.
For the low-viscosity Newtonian case (water), a peak in the drop response was observed around \SI{45}{\hertz}, corresponding to a resonance-like behavior.
In contrast, the high-viscosity Newtonian case (pure glycerin) exhibited a strongly damped response across all frequencies, consistent with overdamped dynamics.
This is in line with the response of a damped driven oscillator.
The polymer solutions were categorized into three regimes:
\begin{itemize}
    \item  Linear viscoelastic regime (small deformations, $\epsilon< 10 $\%).
    \item Nonlinear viscoelastic regime (large deformations, $10 \%<\epsilon< 20 $\%).
    \item Highly nonlinear viscoelastic regime, $\epsilon> 20\%$.
    
    \end{itemize}

In the linear regime, drops with the smallest Deborah numbers (i.e., 3\%, \SI{300}{\kilo\gram\per\mol}) behaved similarly to underdamped Newtonian drops.
At intermediate Deborah numbers (e.g., 3\%, \SI{1000}{\kilo\gram\per\mol}), viscous effects dominated, as reflected in the energy dissipation ($W_{\text{diss}}$) and the time evolution of the drop profile.
For highly elastic cases ($De \gg 1$), elasticity dominated the response, leading to flatter Lissajous curves—an indicator of reduced phase lag between stress and strain.
In the nonlinear regime, achieved by increasing the deformation amplitude, higher harmonics appeared in the drop response and were clearly reflected in both the Lissajous curves and the drop shape evolution.
Overall, this study serves as a proof of concept for drop-based rheology.
The approach enables access to a broader range of frequencies than conventional rheometers, making it a promising technique for characterizing complex fluids, especially when sample volume is limited or high-frequency behavior is of interest.

\begin{acknowledgments}
This study was funded by German Research Foundation (DFG) within the Collaborative Research Centre 1194 “Interaction of Transport and Wetting Processes,” Project- ID 265191195, subprojects A02.
The authors wish to thank Dr.Thomas C. Sykes (University of Warwick) and Dr. José Alberto Rodríguez Agudo ( Anton Paar GmbH) for valuable discussions.
The authors wish to thank Mr. Niklas Gerlach for his contribution to the drawing of the setup schematic.
\end{acknowledgments}

\section*{Data Availability Statement}

Data will be made available upon reasonable request to the corresponding author.

\nocite{*}

\bibliography{apssamp}% Produces the bibliography via BibTeX.

\end{document}